\documentclass[12pt,onecolumn,twoside]{IEEEtran}

\usepackage[noadjust]{cite}
\usepackage[pdftex]{graphicx}
\usepackage[cmex10]{amsmath}%, mathtools}
\usepackage{amsfonts, amssymb, amsthm}
\usepackage{algorithmic}
\usepackage{algorithm}
\usepackage{array}
\usepackage[usenames,dvipsnames]{color}
\usepackage[caption=false,font=footnotesize]{subfig}
\usepackage{setspace}

\theoremstyle{definition}\newtheorem{Rem}{Remark}
\theoremstyle{definition}
\theoremstyle{definition}
\theoremstyle{definition} \newtheorem{Lem}{Lemma}
\theoremstyle{definition}\newtheorem{Thm}{Theorem}
\theoremstyle{definition}\newtheorem{Prop}{Proposition}
\theoremstyle{definition}
\theoremstyle{definition}

\newcommand{\alg}{\bf}

\makeatletter
\renewcommand{\ALG@name}{Procedure}
\makeatother

\allowdisplaybreaks[1]

\begin{document}

\setstretch{1.5}

\title{Complexity Analysis of Heuristic Pulse Interleaving Algorithms for Multi-Target Tracking with Multiple Simultaneous Receive Beams}

\author{Dae-Sung~Jang
        and Han-Lim~Choi*% <-this % stops a space
\thanks{D.S. Jang and H.-L. Choi are with the Division of Aerospace Engineering, KAIST, Daejeon, 305-701, Republic of Korea. E-mail: (dsjang@lics.kaist.ac.kr, hanlimc@kaist.ac.kr).}% <-this % stops a space
%\thanks{Manuscript received April 19, 2005; revised January 11, 2007.}
\thanks{*All correspondence should be forwarded to H.-L. Choi; Mailing address: 291 Daehak-ro, Rm. E4-C327, C-FRIEND Field Robotics Center, KAIST, Yuseong, Daejeon 305-701, Rep. of Korea.; Tel:+82-42-350-3727; E-mail: hanlimc@kaist.ac.kr}
}

\maketitle

%\vspace*{.3in}

\begin{abstract}
This paper presents heuristic algorithms for interleaved pulse scheduling problems on multi-target tracking in pulse Doppler phased array radars that can process multiple simultaneous received beams.
The interleaved pulse scheduling problems for element and subarray level digital beamforming architectures are formulated as the same integer program and the asymptotic time complexities of the algorithms are analyzed.
\end{abstract}


\IEEEpeerreviewmaketitle

%\newpage

\section{Introduction}\label{sec:intro}

%{\red A multi-function radar is designed to fulfill multiple radar functions each of which is previously assigned to a dedicated single function radar.}

In target detection with a radar, round-trip electromagnetic wave propagation between the radar and the target forms an intervening idle time between a transmitted pulse (T-pulse) and a received pulse (R-pulse).
If the intervening idle time can instead be effectively used for some other radar tasks, the utilization level of the radar resource will be increased, potentially leading to enhanced overall radar performance.
A technique that inserts T-/R-pulses in the idle time of other pulses is called \textit{pulse interleaving}.
The pulse interleaving technique utilizes hidden resources of a radar, i.e., unused time.
Thus, for example, pulse interleaving of tracking tasks reduces the total tracking time for a given number of targets, and the saved time can be used for tracking additional targets or enhancing the radar search performance.

There are two types of pulse interleaving: task-level interleaving and pulse-level interleaving.
In task-level interleaving, a radar task, which may consist of several looks or a look with some number of pulses is the unit of interleaving.
Transmitted and received dwells of a task are interleaved with the dwells of other tasks and thus a typical target range is required to be long for a sufficient idle time compared with the dwells.
Meanwhile, \textit{pulse-level interleaving} is available if a group of pulses in a look has sufficient intervening idle times between the pulses.
This level of interleaving allows R-pulses to be interleaved not only with the pulses of other looks but with the T-pulses from which the R-pulses originate.
Therefore, a longer dwell time of a look can be used for a comparatively shorter target range.

Some previous approaches have addressed pulse interleaving of radar tasks \cite{Far80,Orm96,Shi03,Lee06,Gho06,Els04,Tin09}, but most of them dealt with task-level interleaving \cite{Orm96,Shi03,Lee06,Gho06,Els04}.
The previous studies of task-level pulse interleaving assumed that the lengths of intervening idle times were exactly known or well predicted.
Minimizing the maximum completion time of tasks is an NP-hard problem even in this case \cite{Gup96,Orm97,Age07}, and thus heuristic algorithms \cite{Gup96} and approximation algorithms \cite{Els04,Age07} were presented for the problem.
In pulse-level interleaving, additional considerations are required regarding radar system capabilities and interleaving constraints brought from signal processing requirements;
any extensions from the studies of task-level interleaving are not straightforward and have not been investigated.
%and {\red may cause the lack of sufficient previous researches}.

The task-level interleaving in a multi-function phased array radar is usually formed as a coupled-task scheduling problem \cite{Orm96,Orm98}, which decides on the time and the order of performing the tasks;
each coupled-task consists of a pair of T-/R-pulses and an intervening idle time that allows the task scheduler to interleave other coupled-tasks into it.
\iffalse
A heuristic for coupled-task scheduling of a multi-function phased array radar was presented in .
\fi
Pulse interleaving enhances utilization of the radar resource by reducing unused time, whereas the radar consumes more energy and produces more heat.
Therefore, some constraints that reflect limitations of energy resources and heat generation were considered in the pulse interleaving of coupled-tasks \cite{Shi03,Lee06,Gho06}.
In \cite{Shi03}, an energy constraint was considered in real-time template based coupled-task scheduling.
The concept of a schedulability envelope was introduced in \cite{Lee06} to provide a quantitative abstraction for a schedulability check under duty cycle and energy constraints;
multi-target tracking was controlled by an on-line scheduler with the schedulability envelope to enhance radar system utilization while satisfying these constraints.
In \cite{Gho06}, radar power constraints were considered in near-optimal resource allocation by a quality of service (QoS) optimization and the scheduling of coupled-tasks.

The results of task-level interleaving have not been extended to pulse-level interleaving, since pulse repetition in a look, which restricts the feasibility of pulse interleaving, must be considered in pulse-level interleaving.
In pulse-level interleaving, which is usually applied to pulse Doppler radars, the pulses are repeated with a specified frequency for signal processing requirements.
The frequency of these pulses is called the \textit{pulse repetition frequency} (PRF) and its inverse is the pulse repetition interval (PRI).
The PRF/PRI can vary among tasks since available PRFs for target detection are dependent on eclipsing, clutter conditions, and the range and radial velocity of the target.
Therefore, PRF selection of interleaved looks in pulse-level interleaving is a significant factor as looks of different PRFs cannot be mutually interleaved in order to prevent overlapping in the time axis.
This makes direct extensions of the previous task-level works to pulse-level interleaving difficult.

In pulse-level interleaving under the pulse Doppler scheme, a radar has to alternate waveforms and beam directions from pulse to pulse at every PRI.
A phased array radar (PAR) system enables this pulse-to-pulse alteration by rapid electronic beam steering.
The pulse interleaving problem in a pulse Doppler PAR was first considered in \cite{Far80} for multi-target tracking and a heuristic algorithm was presented to reduce the tracking time of targets.
Several heuristic rules were also studied in \cite{Tin09} with energy constraints and priorities of tasks, and all the tasks were assumed to use the same PRF.
However, in both studies, the authors did not address how to determine the PRFs of interleaved looks from different targets, each of which has a different set of available PRFs.
%The absence of PRF selection in the literature

In previous works \cite{Far80,Tin09}, it was assumed that R-pulses as well as T-pulses must not overlap each other since signals from different directions cannot be separated without an appropriate beamforming technique.
A digital beamforming (DBF) technique of a PAR can identify the signals from different directions and enhance interleaving possibilities by relieving one of the constraints of interleaving pulses.
By DBF, a receive beam can be digitally re-steered and processed to form multiple simultaneous receive beams \cite{Sko07};
in other words, a received signal from a wide beam is processed as multiple signals from different directions at the same time.
It has been known that for search operations, multiple simultaneous receive beams (MSRB) bring a reduction of search frame time \cite{Moo97,Sko07}.
Also, for pulse interleaving in multi-target tracking, the MSRB technique allows R-pulses to overlap each other, and thus more pulses can be interleaved and the overall track occupancy is reduced.
The surplus radar time by the reduced track occupancy can be utilized for tracking more targets or can be distributed to other radar functions to enhance their performance.

In this paper, pulse-level interleaving with MSRB by using DBF is considered for multi-target tracking with a pulse Doppler PAR, where DBF is implemented in an element or a subarray level.
While the element level DBF (EDBF) case provides physical insights for pulse interleaving with a relatively idealized setting, the subarray level DBF (SDBF) case applicable to most PARs in service represents a more practical situation.
For these two different levels of DBF architecture, the interleaving problems are formulated as the same integer program.
In EDBF, a received beam can be re-steered in an arbitrary direction, whereas the re-steering direction is limited in SDBF.
Therefore, the pulse interleaving using MSRB in SDBF has, additionally, a selection problem of targets that will be tracked by a group of interleaved looks within the region of the re-steerable direction.
For the problems of both DBF levels, heuristic pulse interleaving algorithms are presented to produce feasible solutions in a computationally tractable fashion.
%The practicality of the algorithms, in terms of performance and computation time against the integer programming, is validated by algorithm complexity analysis and numerical simulations.

The contribution of this paper is threefold.
First, pulse-level interleaving with MSRB by using digital re-steering is, to the authors' knowledge, considered for the first time for multiple target tracking in a pulse Doppler PAR.
Second, this paper also includes considerations of PRF selection for pulse interleaving in both DBF levels and a geometric constraint of MSRB in SDBF.
Third, heuristic pulse interleaving algorithms are presented as practical solvers for both problems of the two DBF levels, and their practicality is validated by an algorithm complexity analysis.

The rest of this paper is organized as follows.
Section \ref{sec:bg} describes physical constraints and properties related to  pulse interleaving for a pulse Doppler PAR with MSRB of the two different DBF levels.
In section \ref{sec:pf}, the mathematical formulation of the pulse interleaving problems for multiple target tracking are presented.
In section \ref{sec:alg}, the heuristic interleaving algorithms are proposed and the complexity analysis of the algorithms is demonstrated in section \ref{sec:alg_complex},
and section \ref{sec:con} concludes the paper.

\section{Background of Pulse Interleaving}\label{sec:bg}

\subsection{Pulse Interleaving in Pulse Doppler Radar}\label{sec:bg_pdr}

In pulse Doppler radars, a number of pulses in a \textit{look} are transmitted with a certain PRF to acquire target's radial velocity by measuring the Doppler shift of a received signal.
\iftrue
A look is a group of pulses repeated with a certain PRF and it can be regarded as a unit of creating a measurement of a target.
\else
A look is a group of the pulses, which can be regarded as a unit of creating a measurement of a target.
\fi
%For airborne case, a look with a high PRF can measure the Doppler shift unambiguously, but range measurements are highly folded \cite{Den07}.
%On the contrary, a low PRF has good range unambiguity and can reduce sidelobe detection by sensitivity time control, whereas it has poor ground moving target rejection and Doppler ambiguity \cite{Sko07,Den07,Sko01}.
%A medium PRF, usually 10-40kHz, has both range and Doppler ambiguity, but it exhibits compromised performance in all aspect target detection in the presence of clutter \cite{Sko07,Hov82,Wil06}.
The range and the radial velocity of a target are measurable without ambiguity if the range and the Doppler shift are smaller than an unambiguous range $R_u$ and an unambiguous frequency $f_u$, respectively:
\begin{equation}\label{e1}
        R_{u}=\frac{ct_{r}}{2}=\frac{c}{2f_{r}},\;\;f_{u}=f_{r}
\end{equation}
where $c$ is speed of light, $t_r$ is PRI, and $f_r$ is PRF.
The received signal from the target is folded when the actual target range is greater than $R_u$ or the Doppler frequency is larger than $f_u$.
An ambiguous range $R_a$ and an ambiguous frequency $f_a$ are measurable quantities from the folded received signal:
\begin{equation}
        R_{a}\equiv R\mod\frac{c}{2f_{r}}
\end{equation}
\begin{equation}
        f_{a}\equiv\frac{-2V_{t}}{\lambda}\mod f_{r}\equiv f_{s}\mod f_{r}
\end{equation}
where $R$ is the range between the target and the radar, $V_t$ is the target's radial velocity with respect to the ground, $\lambda$ is the wave length of the radar signal, and $f_s$ is the Doppler shift of the received signal.

In a search beam, multiple PRFs are used to resolve the range and Doppler ambiguities since the radar system has no prior information of a detected target.
However, a track beam for a moderate speed target can be formed of single-PRF looks using the target's range/velocity estimate and covariance from a tracking filter.
Thus, in this paper, a tracking task of a target uses a single PRF at each track update, selected among a predesigned set of PRFs.

\iffalse
Under the assumptions of blind zones above, a central PRF value that gives the maximum visibility of a target in both range and frequency domain is 13kHz \cite{Sko07}.
The central PRF is a standard point for composing a PRF set to make the range and frequency domain of interest clear.
The number PRFs in the set is typically 8 to minimize ghost detection and to resolve ambiguities of range and frequency domain \cite{Wil06}.
The set is designed to provide the visibility more than 96\% and it must be clear at maximum design range \cite{Sko07}.
Fig. \ref{fig2} shows a blind map of 8 PRFs (9.5, 10.5, 11.5, 12.5, 13.5, 14.5, 15.5, 16.5kHz) which satisfies the criteria.
This set of PRFs is used in the simulation.
\fi
% For medium PRF airborne radar, sidelobe clutter covers entire frequency domain, thus the return of target is always in sidelobe clutter region.
% If signal to noise ratio (SNR) is sufficient to detect the target

\begin{figure}[t]
\centerline{
    \includegraphics[width=.8\columnwidth]{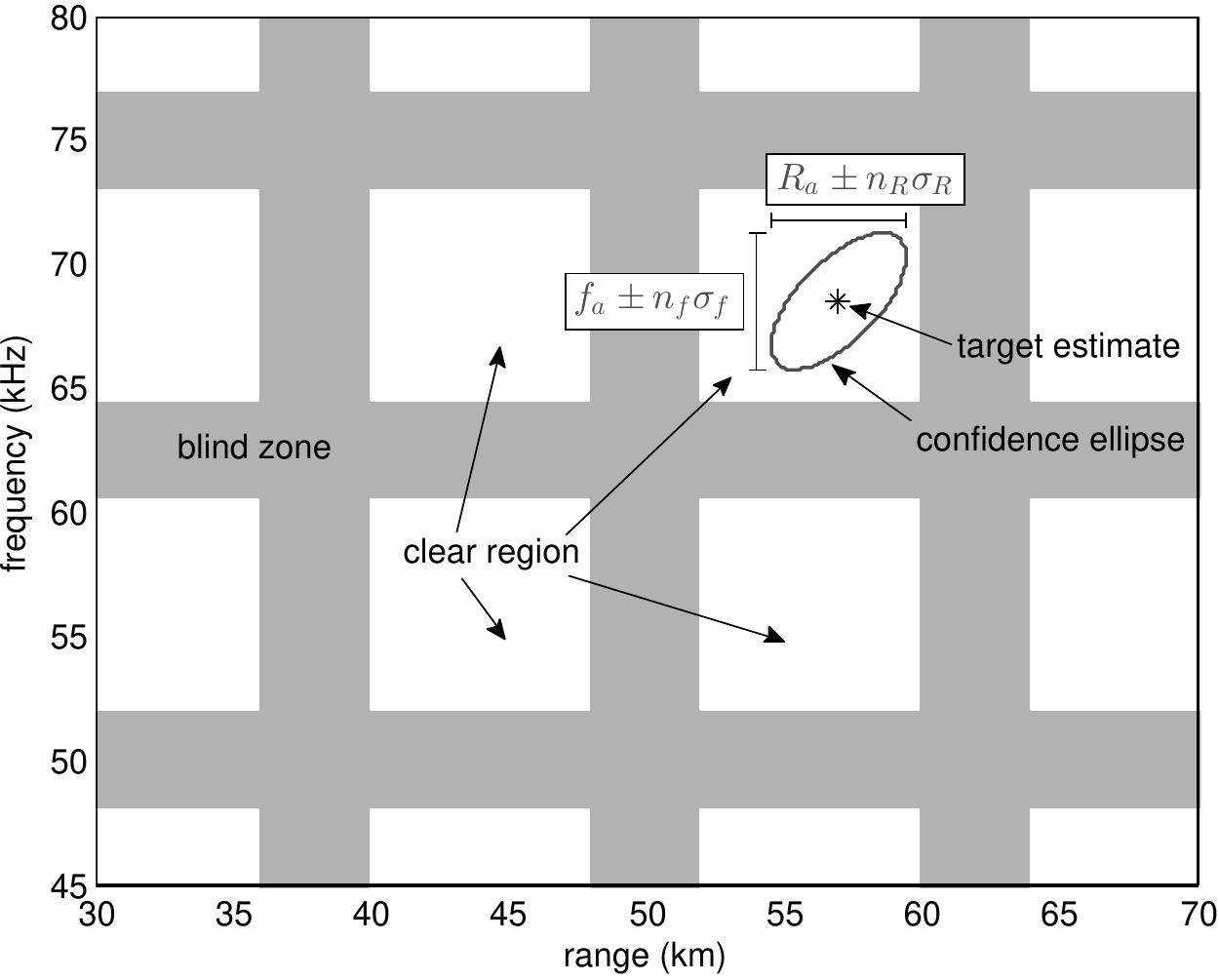}}
    \vspace*{-.1in}
    \caption{The blind zones (gray) and clear regions (white) of target tracking with $f_r$=12.5kHz. The ellipse depicts a boundary specified by multiples of standard deviations for the estimate(*).}
    \label{fig1}
    \vspace*{-.2in}
\end{figure}

Target detection is greatly restricted in blind zones induced by the losses from pulse eclipsing and clutter in the range and frequency domain of interest.
In this paper, it is assumed that the blind zones are placed around the folded frequencies and ranges that are the integer multiples of $f_u$ and $R_u$, respectively.
This can be appropriate when a rejection notch is at main beam clutter and a target's radial velocity with respect to the ground is measured in a Doppler pass band \cite{Sko07,Hov82,Lon85}.
Fig. \ref{fig1} shows the blind zones and the clear regions of $f_{r}$=12.5kHz as a sample of medium PRF for an airborne radar;
the widths of the blind zones are assumed to be 4kHz for frequency and 4km for range.
As an example, the estimated range and the Doppler shift (equivalently, the radial velocity) of a target obtained from some tracking filter is plotted in Fig. \ref{fig1} with a confidence ellipse specified by sigma multiples $n_R$ and $n_f$ of standard deviations of the range and the Doppler shift.
If this confidence ellipse lies totally inside the clear region, the target is highly likely to be detected at the next track update.

The confidence intervals of the target range and the Doppler frequency, which contain the confidence ellipse, are described as $R_a \pm n_R \sigma_R$ and $f_a \pm n_f \sigma_f$ with appropriately chosen multiples $n_R$ and $n_f$, where $\sigma_R$ and $\sigma_f$ denote standard deviations of the target range and Doppler frequency.
Then, if the confidence intervals lie completely within the clear region of the associated PRF, the target can be detected using the PRF with a probability, at least, specified by the choice of $n_R$ and $n_f$.
Therefore, the following conditions need to be satisfied to ensure reliable tracking of the target with $f_r$:
\begin{eqnarray}
       & R_{a}-n_{R}\sigma_{R} \geq \epsilon_{R}^+ , &R_{a}+n_{R}\sigma_{R}\leq R_{u}-\epsilon_{R}^- \label{e4} \\
       & f_{a}-n_{f}\sigma_{f} \geq \epsilon_{f}^+, & f_{a}+n_{f}\sigma_{f}\leq f_{u}-\epsilon_{f}^-. \label{e5}
\end{eqnarray}
where left-/right-end blind widths in range direction, $\epsilon_R^+$ and $\epsilon_R^-$, and frequency direction, $\epsilon_f^+$ and $\epsilon_f^-$ are defined as
%with the left-end blind widths $\epsilon_{R}^+, \epsilon_{f}^+$ and right-end blind widths $\epsilon_{R}^-, \epsilon_{f}^-$ are defined below.
\begin{eqnarray}
        &\epsilon_{R}^+=\max \left\{C_{R}^+,\frac{ct_{p}}{2} \right\}, &\epsilon_{R}^-=C_{R}^-+\frac{ct_{p}}{2}, \label{e6} \\
        &\epsilon_{f}^+=C_{f}^+, &\epsilon_{f}^-=C_{f}^- \label{e7}
\end{eqnarray}
where $C_{R}^+$, $C_{R}^-$ denotes left-/right-end clutter regions along the range domain, $C_{f}^+$, $C_{f}^-$ are left-/right-end clutter regions along the frequency domain, and $t_p$ represents the pulse width of a track beam.
%Hereafter, if a target estimate satisfies the conditions in (\ref{e4}) and (\ref{e5}) for a certain PRF, then the target and the PRF are called \textit{compatible} to each other.

If a target can be tracked with some PRF and there are enough intervening idle times of T-/R-pulse trains in a tracking task of the target, it is possible to interleave the pulse trains of another target's task with the same PRF.
On the other hand, if the PRFs of two looks are different, it is highly probable that pulses in the looks are overlapped, in particular, when the looks consist of a large number of pulses.

\begin{figure}[t]
\centerline{
    \includegraphics[width=.9\columnwidth]{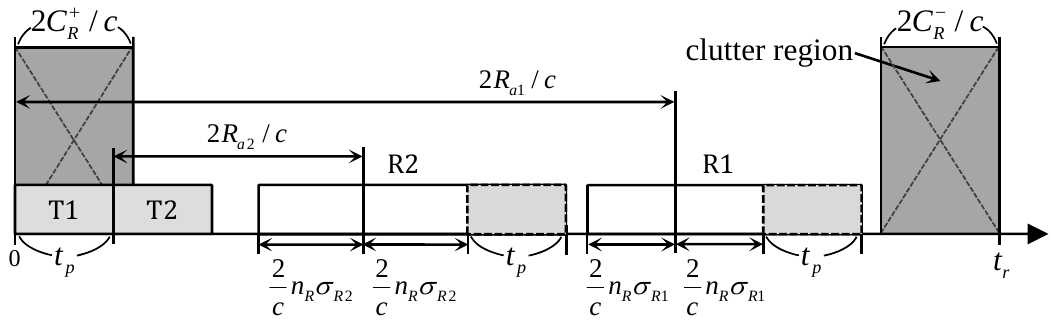}}
    \vspace*{-.1in}
    \caption{An illustrative pulse interleaving scheme in a PRI for a pulse Doppler PAR without MSRB (i.e., with the non-overlapping constraint for received beams). T-/R-pulses of tasks are marked with `T' and `R' followed by the task numbers. $R_{a1}$ and $\sigma_{R1}$ are the range and the standard deviation of the target of task 1, and $R_{a2}$ and $\sigma_{R2}$ are the ones of task 2.}
    \label{fig2a}
    \vspace*{-.15in}
\end{figure}

For example, consider three targets that are to be tracked with three different PRFs:
suppose that target 1 is trackable with PRF 1 and 2; target 2 is trackable with PRF 2 and 3; target 3 is trackable with PRF 3 only. %and target 4 is trackable with PRF 1 and 3.
Also, assume that each target is assigned to a tracking task consisting of several looks with an available PRF, and the periods of tracking tasks are similar and the requested intervals of task execution are densely distributed over a short time.
Then, the pulses of tasks trackable with the same available PRF can be interleaved; one possible pulse interleaving scenario for PRF 2 is to interleave pulse trains for target 1 and 2.
%in this particular example, interleaving is not possible for other PRFs.
Fig. \ref{fig2a} illustrates an example of pulse interleaving in a PRI for the tasks of the target 1 and 2.
As shown in the figure, all T-pulses and possible intervals of R-pulses defined by the range estimates and standard deviations should not overlap with each other at any time points.
Also, note that as a T-pulse is delayed by other preceding T-pulses, the corresponding R-pulse is delayed by the same amount of time.

\subsection{Pulse Interleaving with Multiple Simultaneous Receive Beams}\label{sec:bg_msrb}

\begin{figure}[t]
\centerline{
    \includegraphics[width=.9\columnwidth]{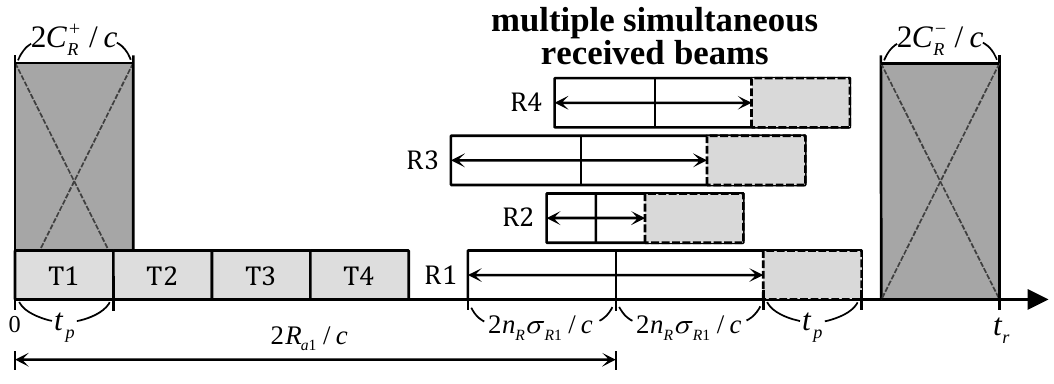}}
    \vspace*{-.1in}
    \caption{An illustrative  pulse interleaving scheme in a PRI for a pulse Doppler PAR with MSRB; more tasks can be interleaved compared to Fig. \ref{fig2a}. T-/R-pulses of tasks are marked with `T' and `R' followed by the task numbers. $R_{a1}$ and $\sigma_{R1}$ are the range and the standard deviation of the target of task 1.}
    \label{fig2b}
    \vspace*{-.25in}
\end{figure}

In pulse interleaving for multiple target tracking, the non-overlapping constraint of R-pulses can be relieved by the MSRB technique \cite{Sko07,Ric10,Mel12}.
The MSRB technique is implemented by using DBF and re-steering of received beams;
in multiple target tracking, a radar can handle multi-target returns from different directions even if the pulses are actually received simultaneously.
Fig. \ref{fig2b} depicts an illustrative pulse interleaving scheme with MSRB in a PRI. %, in case four targets are trackable with the associated PRF.
Observe that, with MSRB, more tasks can be interleaved with each other compared to the case without it (Figs. \ref{fig2a} and \ref{fig2b} are scaled equally for comparison).

Pulse Doppler PARs using solid state transmit/receive modules typically have lower peak power and a higher duty ratio;
thus, the maximum number of interleaved pulses in a PRI is limited to some small number.
Furthermore, the estimated time intervals of received signals have much longer lengths than T-pulses.
Therefore, for pulse interleaving schemes in this work, it is assumed that T-pulses are successively positioned from the starting time of each PRI followed by R-pulses as seen in Fig. \ref{fig2b}.
This assumption is not simplistic and represents most physically feasible cases in that interleaving R-pulses, which can be processed simultaneously, between T-pulses in a PRI is rarely beneficial.
It is also assumed that the interleaving of successively attached pulses satisfies radar duty ratio and does not violate any energy consumption constraints of the radar.

Since interleaved T-pulses of tracking tasks are transmitted separately, any adequate beamforming for individual tracks to different directions is allowed on transmission.
On the other hand, the directions of received beams are restricted by re-steering capability of a PAR, where the capability is determined by the type of the DBF architecture.
In EDBF, received signals are digitized at each array element whereas a combined received signal at each subarray is digitized in SDBF.
The digitized signal is weighted and delayed in a digital beamformer where multiple independent beams can be formed in different directions.
Although EDBF allows independent beams to be steered into any direction~\cite{Sko07}, SDBF is preferably implemented for practical reasons such as size, cost, and computing power.
In SDBF, only the array factor of a received beam can be digitally re-steered, but the subarray pattern of the received beam is maintained at an originally steered direction set by phase shifters.
Hence, grating lobes of the array factor escaped from the nulls of the subarray pattern create high sidelobes \cite{Ric10}.
This can be attenuated by an overlapped subarray and low sidelobe weighting, but the re-steering direction is limited: the extent is within the subarray pattern \cite{Sko07,Ric10,Mel12}.

\begin{figure}[t]
\centerline{
    \includegraphics[width=0.75\columnwidth]{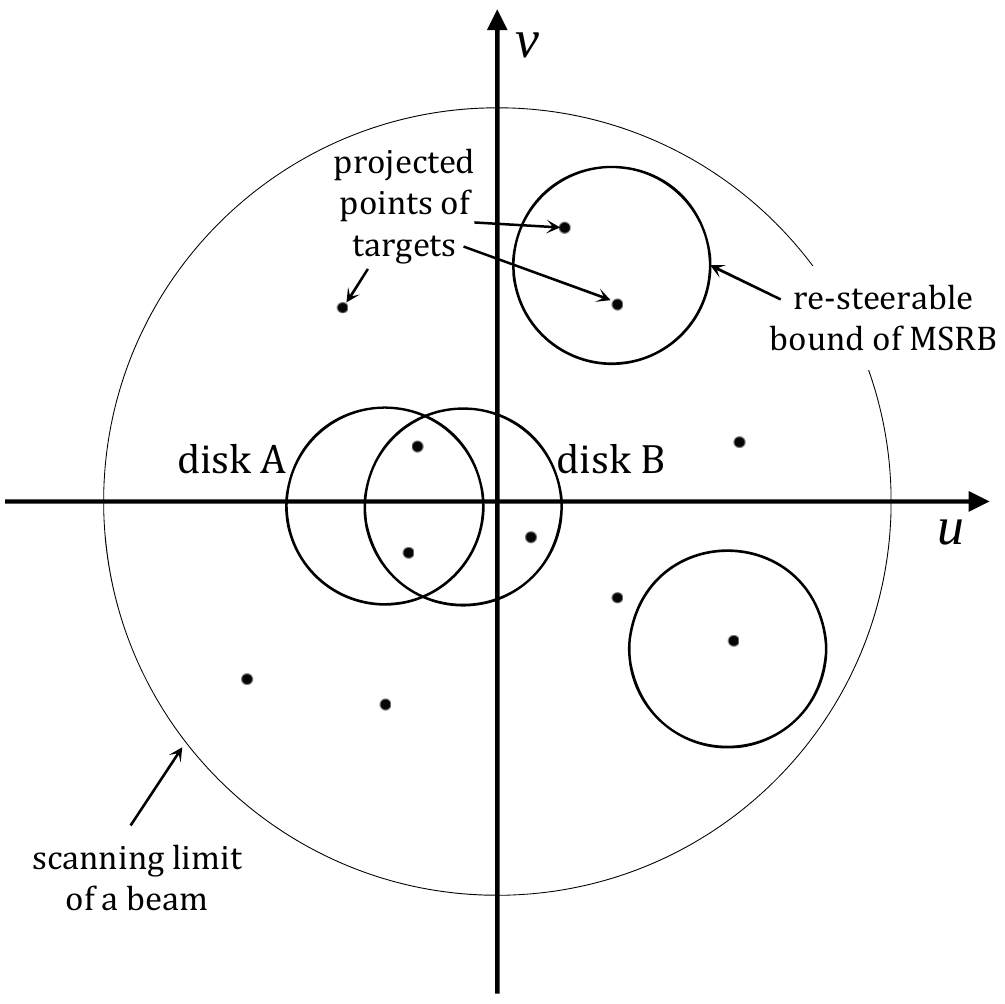}}
    \vspace*{-.13in}
    \caption{Projected points of targets trackable with a PRF and circular re-steering boundaries of MSRB, i.e., the disks on the normalized scanning plane. $u$ and $v$ are the horizontal and vertical domains of the plane.}
    \label{fig3}
    \vspace*{-.25in}
\end{figure}

Due to the limitation in re-steering capability of MSRB, pulse interleaving with SDBF needs to satisfy some angular nearness condition of multiple targets in addition to the PRF availability described in section \ref{sec:bg_pdr}.
The interleaving possibility of targets under the limited re-steering can be discriminated in a plane of direction cosines of beam scanning direction, called T plane in \cite{Aul60}.
In this paper, the \textit{normalized scanning plane} denotes the plane of direction cosines, as any point on the plane is a normalized projection from a point on a hemisphere where a target is located.
The normalized scanning plane is confined in a unit disk centered at the origin.
Fig. \ref{fig3} depicts projected points of targets trackable with the same PRF and re-steering boundaries of MSRB on the normalized scanning plane.
Since trackable targets are different depending on the PRF, a similar figure can be drawn for each PRF.
If a circle beam is used and the corresponding subarray pattern is also circular, the re-steerable area of MSRB forms a disk.
Some examples of such disks are plotted in Fig. \ref{fig3}: targets within each disk are allowed to be interleaved together.
Without considering directivity loss and other factors, a disk with more targets is preferred as long as the targets are enclosed by the disk.
In Fig. \ref{fig3}, disk A encloses two targets around the origin and disk B encloses another target in addition to the targets, thus it is enough to consider disk B for pulse interleaving.
In this way, the limitation of the pulse interleaving using MSRB in SDBF can be identified as this geometric constraint of equal-sized disks covering projected target points on the normalized scanning plane for each PRF.

\section{Problem Formulation}\label{sec:pf}

Optimization problems to reduce tracking time in pulse interleaving with MSRB of both EDBF and SDBF are formulated in this section.
The problem under EDBF is firstly considered as an integer program.
The problem at SDBF is also represented by the same integer program, but groups of tasks that can be interleaved together need to be identified in advance.

For both problems, each target is assumed to generate a single tracking task that consists of several looks with a certain PRF.
If the periods of tracking tasks are similar and the requested intervals of task execution are densely distributed over a short time, at the time of tracking updates of all targets, pulses of some tasks can be interleaved with the same PRF.
For notational convenience, hereafter, an interleaved group of looks is called simply an \textit{interleaved look} or a look, regardless of the number of looks contained in a single task.
In each PRI of a look, multiple consecutive time slots of T-pulses can be specified since the T-pulses of different tasks are assumed to be successively positioned (see Fig. \ref{fig4}).
If a T-pulse of a task is located in one of the slots in the first PRI, all the same slots in all later PRIs of the look is also occupied by the T-pulses of the same task since the pulses of the tasks are repeated with the same PRF in the interleaved look.

\subsection{Pulse Interleaving with EDBF}\label{sec:pf_edbf}

\begin{figure}[t]
\centerline{
    \includegraphics[width=.9\columnwidth]{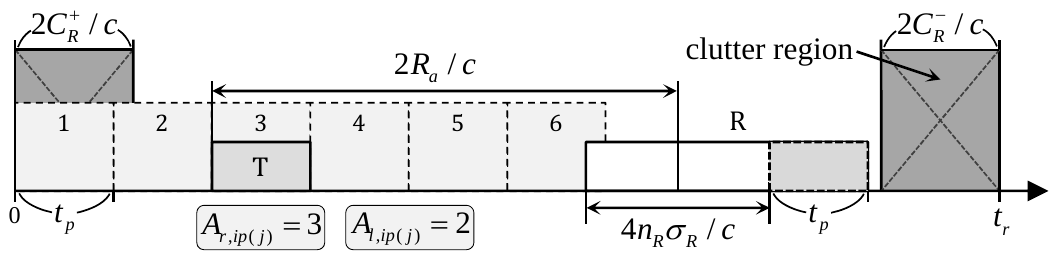}}
    \vspace*{-.05in}
    \caption{The leftward availability $A_{l,ip(j)}$ and the rightward availability $A_{r,ip(j)}$ of the task $i$ for the $j$th look with a PRF $p$. Two T-pulses of other tasks can be interleaved between T-/R-pulses of the task $i$ ($A_{l,ip(j)}$=2), and the T-pulse of the task $i$ can be shifted up to the third slot ($A_{r,ip(j)}$=3). The consecutive possible slots of T-pulses are expressed as light gray squares with dotted lines and marked with the slot numbers.}
    \label{fig4}
    \vspace*{-.2in}
\end{figure}

In the pulse interleaving problem with EDBF, for the $i$th target ($\equiv$$i$th task) and the $j$th look with a PRF $p$, $A_{v,ip(j)}$ indicates availability of the task-look pair $(i,j)$.
$A_{v,ip(j)}$ is 1 if the target $i$ satisfies (\ref{e4}) and (\ref{e5}) with the $j$th look's PRF, otherwise $A_{v,ip(j)}$ is 0:
\begin{equation}
        A_{v,ip(j)}=
        \begin{cases}
            1, & \text{if $(i,j)$ pair satisfies (\ref{e4}) and (\ref{e5}),}\\
            0, & \text{otherwise}.
        \end{cases}
\end{equation}
The leftward availability $A_{l,ip(j)}$ represents the number of tasks whose T-pulses can be interleaved between T-/R-pulses of task $i$ (see Fig. \ref{fig4}):
\begin{equation}\label{e9}
        A_{l,ip(j)}=A_{v,ip(j)}\cdot\max\left(0,\left\lfloor\frac{2}{ct_{p}}(R_{a,ij}-n_{R}\sigma_{R,i}-\epsilon_{R}^+)\right\rfloor\right),
\end{equation}
where $R_{a,ij}$ signifies the ambiguous range of the target $i$ with the PRF of the $j$th look, and $\sigma_{R,i}$ is the standard deviation of the range of the target $i$.
The rightward availability $A_{r,ip(j)}$ indicates the number of slots counting rightward that the T-pulses of the task $i$ can be shifted while keeping the task's R-pulses within the clear region (see Fig. \ref{fig4}):
\begin{IEEEeqnarray}{lll}\label{e10}
        A_{r,ip(j)}={}&A_{v,ip(j)}\\\nonumber
        &\cdot\max\!\left(\!0,\!\left\lfloor\frac{2}{ct_{p}}(R_{u}\!-\!(R_{a,ij}\!+\!n_{R}\sigma_{R,i}\!+\!\epsilon_{R}^-))\!+\!1\right\rfloor\right).
\end{IEEEeqnarray}

The goal of the pulse interleaving is to minimize the total tracking time, i.e., the sum of the dwell times for tracking all the targets, by optimally assigning the tracking tasks to the looks.
There are two binary decision variables:
$h_{ijk}$ is 1 when the T-pulse of the $i$th tracking task is scheduled at the $k$th slot of the $j$th look, and is 0 otherwise;
$f_{j}$ is 1 if there is at least one task assigned to the $j$th look.
With the decision variables, the pulse interleaving with MSRB for the EDBF case can be formulated as an integer program:
%\begin{IEEEeqnarray}{ll}
\begin{align}
        \underset{h_{ijk}, f_{j}}{\text{minimize}} \;\;&\sum_{j}^{N_l}t_{d,j}f_{j}\label{e0f1} \tag{\textbf{IP}} \\
        \text{subject to} \;\;&\sum_{i}^{N_t}\sum_{k}^{N_{\text{intlv}}}h_{ijk}\leq N_{\text{intlv}}f_j\label{e0f2} \tag{C1}\\
        &\sum_{j}^{N_l}\sum_{k}^{N_{\text{intlv}}}h_{ijk}=1\label{e0f3} \tag{C2}\\
        &\sum_{i}^{N_t}h_{ijk}\leq1\label{e0f4} \tag{C3} \\
        &\sum_{i}^{N_t}(h_{ijk}-h_{ij(k+1)})\geq0\label{e0f5} \tag{C4} \\
        &\sum_{k}^{N_{\text{intlv}}}h_{ijk}\leq A_{v,ip(j)}\label{e0f6} \tag{C5} \\
        &\sum_{i}^{N_t}k\cdot h_{ijk}\leq\sum_{i}^{N_t}A_{r,ip(j)}h_{ijk}\label{e0f7} \tag{C6}\\
        \sum_{i}^{N_t}\!\sum_{k}^{N_{\text{intlv}}}\!h_{ijk}&\leq\!\sum_{i}^{N_t}(k\!+\!A_{l,ip(j)})h_{ijk}\!+\!L_{\infty}(1\!-\!\sum_{i}^{N_t}h_{ijk})\label{e0f8} \tag{C7} \\
        &h_{ijk}\in\{0,1\},\;f_{j}\in\{0,1\}\label{e0f9} \tag{C8}
\end{align}
%\end{IEEEeqnarray}
% &\sum_{i}\sum_{j}(h_{ijk}t_{p})\leq\sum_{j}h_{ijk}\biggl\{(j-1)t_{p} \nonumber\\
% &\;\;\;\left.+{2(R_{i}-n_{R}\sigma_{R_i})\over c}\right\}+L_{\infty}(1-\sum_{j}h_{ijk})\label{e0f6}\\
% &h_{ijk}=0 \text{ for specified (i,k) pairs and all }j\text{s}\nonumber
where $t_{d,j}$ is the dwell time of the $j$th look,
$N_{\text{intlv}}$ is the maximum number of tasks that can be interleaved in a single look,
$L_{\infty}$ is an arbitrary large number, and
$N_{l}$ and $N_t$ are the total number of looks and the targets (equivalently tasks), respectively.
It should be noted that $N_l$ is typically large -- as many as $N_t N_{\text{PRF}}$ with $N_{\text{PRF}}$ being the number of possible PRFs in a radar system -- since a sufficient number of looks need to be specified for each PRF in order for pulse interleaving to be exact.

Also, following constraints are involved in the integer program (\ref{e0f1}).
(\ref{e0f2}) the number of tasks that can be interleaved in a single look is limited by $N_{\text{intlv}}$;
(\ref{e0f3}) a task must be scheduled only one time;
(\ref{e0f4}) a slot of a look can be designated to at most a single task;
(\ref{e0f5}) interleaved T-pulses in a PRI are placed successively;
(\ref{e0f6}) a task can be interleaved in a look with an available PRF of the task;
(\ref{e0f7}) the $k$th slot of a look can be occupied by a T-pulse of a task whose $A_{r,ip(j)}$ is larger than or equal to $k$;
(\ref{e0f8}) all T-pulses of tasks interleaved in a look must not be overlapped with R-pulses of the tasks.

%\subsection{The Relation to The Single-Source Capacitated Facility Location Problem}\label{sec:pf_sdbf}
\begin{Rem} \label{rem:sscfl}
Note that by excluding pulse scheduling in a look and by assuming any target has the maximum left-/rightward availabilities for available PRFs, the integer program (\ref{e0f1}) is simplified as:
%\begin{IEEEeqnarray}{ll}
\begin{align}
        \underset{h_{ij}, f_{j}}{\text{minimize}} \;\;&\sum_{j}^{N_l}t_{d,j}f_j\label{e0s1} \tag{\textbf{IP$_\text{SSCFL}$}}\\
        \text{subject to} \;\;&\sum_{i}^{N_t}h_{ij}\leq N_{\text{intlv}}f_j\label{e0s2} \\
        &\sum_{j}^{N_l}h_{ij}=1\label{e0s3}\\
        &h_{ij}\leq A_{v,ip(j)}\label{e0s4}\\
        &h_{ij}\in\{0,1\},\;f_{j}\in\{0,1\}\label{e0s5}
\end{align}
%\end{IEEEeqnarray}
This type of the problem is called the single-source capacitated facility location problem (SSCFL)~\cite{Kor00}, which is known to be NP-hard~\cite{Nee83,Hol99}. The pulse interleaving formulation in (\ref{e0f1}) includes a more complicated set of constraints than SSCFL; thus, obtaining the optimal solution to (\ref{e0f1}) with a generic integer programming solver may not be scalable.\footnote{This does not exclude possible existence of a tailored exact algorithm for the formulation.}
%There are some previous works about exact solutions \cite{Nee83,Hol99} and heuristics \cite{Sri93,Ron99} for this NP-hard problem.
%A recent study \cite{Ahu04} verifies local search algorithms using multi-exchange heuristic are effective for this problem.
\end{Rem}

\subsection{Pulse Interleaving with SDBF}\label{sec:pf_sdbf}

For pulse interleaving with SDBF, the geometric constraint in the normalized scanning plane described in section \ref{sec:bg_msrb} needs to be considered as well.
Even in this case, the integer programming formulation in (\ref{e0f1}) can still be used if the set of looks with disks enclosing some projected target points are identified in advance and the associated availability parameters are evaluated:
$A_{v,ip(j)}$ is 1 if the task $i$ satisfies (\ref{e4}) and (\ref{e5}) with the $j$th look's PRF and the projected target point of the task $i$ is located in the disk of the $j$th look, otherwise $A_{v,ip(j)}$ is 0;
due to the change of $A_{v,ip(j)}$, the values of $A_{l,ip(j)}$ and $A_{r,ip(j)}$ are also modified according to (\ref{e9}) and (\ref{e10}).
However, the process of identifying the set of looks for the SDBF case is not trivial and requires some type of {\it unique disk search} process.

In the SDBF case, the pulses of tracking tasks can be interleaved in a look only if the corresponding target points in the normalized scanning plane are enclosed by a disk that represents the re-steerable region of MSRB of the look.
Therefore, to obtain an exact and optimal solution for the integer program, it is a prerequisite that all possible looks, in other words, all possible combinations of targets whose projected points are enclosed by the disks of the looks are searched and identified.
If continuous beam pointing is available in the normalized scanning plane, there are an infinite number of disks that can be candidates to form the looks.
Since the number of combinations of the projected target points are finite, the disks enclosing the points should be identified without redundancy for solving the optimization problem efficiently.
Suppose a disk is a set of enclosed points projected from some targets on the normalized scanning plane.
Then, given projected points of all targets, all possible looks and disks are obtained, without redundancy, from the minimum cardinality collection of all unique equal-sized disks each of which is neither empty nor a proper subset of any disks.
Any algorithm for searching the unique disks among infinite possibilities can reduce unnecessary computation.

This unique disk search is a preprocessing for solving pulse interleaving with SDBF by the integer program (\ref{e0f1}) and it is closely related with the unit disk cover problem \cite{Hoc85,Das12}, which determines the minimum cardinality collection of unit-sized disks among given disks that cover all the points of interest in a two-dimensional plane.
However, to the best of the author's knowledge, any relation of the unique disk search to the unit disk cover has hardly received attention in the literature \cite{Hoc85,Cha86,Cla07,Lia10,Mus10,Das12};
thus, no exact algorithm is available to find out the minimum cardinality collection for the unique disk search process.

Reserving the development of an efficient algorithm for this disk search problem as one of future work items, this paper takes an approximate approach that discretizes the solution space to a finite grid, in other words, disk centers are only allowed to be located at the grid points, and utilizes a brute-force search (detailed in section \ref{sec:alg_sdbf}) among the disks on the grid.
As it is noted in section \ref{sec:bg_msrb}, each PRF has a different set of projected points of trackable targets, and thus this procedure is repeated for each PRF.

\section{Heuristic Algorithms}\label{sec:alg}
%The integer program (\ref{e0f1}) of the pulse interleaving with MSRB includes more complicated constraints than the single-source capacitated facility location problem, which is NP-hard.
In multi-target tracking situation, the ranges and Dopper frequencies of moving targets and thus corresponding available PRFs are vary with time.
Therefore, the pulse interleaving problem needs to be solved at every track update, which is typically less than a few seconds.
This requires a computationally efficient algorithm to solve the integer program (\ref{e0f1});
however, as discussed in Remark \ref{rem:sscfl}, a generic integer programming solver is not scalable and thus may not be able to create solutions in a high-frequency manner.
Therefore, developing computationally efficient heuristic algorithms can be a good option to produce acceptable solutions for the pulse interleaving in practice.
This section presents heuristic interleaving algorithms: for the EDBF case (in section \ref{sec:alg_edbf}) and for the SDBF case (in section \ref{sec:alg_sdbf}),
the time complexities of both algorithms are analyzed in the next section (section \ref{sec:alg_complex}).

\subsection{Heuristic Interleaving Algorithm for EDBF}\label{sec:alg_edbf}

The heuristic algorithm for EDBF consists of two phases: PRF selection of an interleaved look and scheduling of interleaved pulses in the look.
The overall procedure of the algorithm is as follows.
The algorithm repeatedly selects some PRF of a look, where pulse interleaving and scheduling are processed, until all tasks are assigned to looks.
Every time a PRF is selected by a heuristic rule, pulses of some tasks are interleaved and scheduled along the slots in the look.
The scheduling starts from the rightmost (latest) slot and a task is selected by a heuristic rule for each slot while the scheduling slot translates backward (leftward).
The two heuristic rules for the PRF and task selection in the algorithm are priority index based rules and thus the PRF/task with the maximum or minimum priority index are selected.
If the PRF heuristic rule always selects some PRF at least one unscheduled task is trackable with, the correctness of the algorithm is dependent on the finite termination and feasibility of the scheduling procedure in a look;
since tasks are scheduled only if their pulses can be interleaved at each slot (see Procedure \ref{alg2}) and the maximum number of possible iterations of the scheduling procedure in a look is $2N_{\text{intlv}}$ (see Proposition \ref{prop2}), the heuristic algorithm always terminates and produces a feasible solution.
%If the PRF heuristic rule always selects a PRF at least one unscheduled tasks is trackable with and the scheduling in a look terminates with a feasible pulse interleaving in a finite time, then the heuristic algorithm always terminates and produces a feasible solution to the interleaving problem.

\begin{algorithm} [t]
    \caption{HeuristicInterleavingElementDBF}
    \label{alg1}
\begin{algorithmic}[1]
%\STATE Load $A_{v,in}$, $A_{l,in}$, and $A_{r,in}$ for all $(i,n)$ pairs.\label{HIED_load_av}
\STATE $H \leftarrow \phi$
\STATE $T_0 \leftarrow$ the set of all tasks $\equiv \{i|\forall i\}$\label{HIED_T_0}
\STATE $j \leftarrow 0$
\WHILE {$T_0\neq\phi$}
    \STATE $j \leftarrow j+1$
    \STATE $p \leftarrow$ select a PRF by a heuristic rule ({\it Heuristic of PRF selection}).\label{HIED_PRF_sel}
    \STATE $T_p \leftarrow \{i|\forall i\in T_0, A_{v,ip(j)}=1\}$\label{HIED_T_p}
    \STATE $S \leftarrow \phi$
    \STATE $S \leftarrow$ BackwardInterleaving($T_p,S,1,N_{\text{intlv}},p$)\label{HIED_call_BI}
    \STATE $T_0 \leftarrow T_0\!\setminus\!\{i|\forall (i,k)\in S\}$\label{HIED_task_del}
    \STATE $H \leftarrow H\cup\{(i,j,k)|\forall (i,k)\in S\}$\label{HIED_sch}
\ENDWHILE
\RETURN $H$\label{HIED_end}
\end{algorithmic}
\end{algorithm}

\subsubsection{Procedure 1 - HeuristicInterleavingElementDBF}

Procedure \ref{alg1} and \ref{alg2} are pseudocodes for the PRF selection and the pulse scheduling procedures.
The presented heuristic algorithm starts from Procedure \ref{alg1} ({\alg HeuristicInterleavingElementDBF} or {\alg HIED}).
%First, in the pseudocodes, $A_{v,in}$ denotes the availability of a task-PRF pair of task $i$ and a look with PRF $n$, and  $A_{l,in}$ and $A_{r,in}$ signify the leftward and rightward availabilities of the pair.
The interleaving result $H$, empty at first, of the algorithm is a set of tuples consisting of task indices ($i$), look indices ($j$), and interleaved slots ($k$).
$T_0$ denotes the set of all tasks, which is contracted by removing scheduled tasks at the end of every while-loop of {\alg HIED}.
At every iteration of the while-loop, a PRF is selected by a PRF selection heuristic (Line \ref{HIED_PRF_sel}; see the detail about the priority index based rules in section \ref{sec:prf_task_rule}).
After PRF $p$ of look $j$ is selected, all tasks in $T_0$ trackable with $p$ are assigned to $T_p$ in Line \ref{HIED_T_p} and then Procedure \ref{alg2} ({\alg BackwardInterleaving} or {\alg BI}) is called for the pulse scheduling in Line \ref{HIED_call_BI}.
As the result of the pulse scheduling in the look $j$, a schedule $S$ is composed of tasks of interleaved pulses and corresponding slots.
The scheduled tasks are eliminated from $T_0$ (Line \ref{HIED_task_del}) and triplets $(i,j,k)$ of the scheduled tasks are inserted in $H$ (Line \ref{HIED_sch}).
If $T_0$ becomes empty, then {\alg HIED} returns $H$ and terminates (Line \ref{HIED_end}).

\begin{algorithm} [t]
    \caption{BackwardInterleaving($T_a,S,E_l,E_r,p$)}
    \label{alg2}
\begin{algorithmic}[1]
\STATE $tail \leftarrow E_r$
\FOR {$cursor \leftarrow E_r$ to $E_l$ step -1}
    \STATE $T_l \leftarrow \{i|\forall i\in T_a, (tail-cursor)\leq A_{l,ip(j)}\}$\label{BI_T_l}
    \STATE $A_{l,S} \leftarrow$ the leftward availability of schedule $S$\label{BI_A_lS}
    \IF {$T_l\neq\phi$}
        \STATE $T_r \leftarrow \{i|\forall i\in T_l, cursor\leq A_{r,ip(j)}\}$\label{BI_T_r}
        \IF {$T_r\neq\phi$}
            \STATE $S \leftarrow$ schedule a task at $cursor$ among the tasks in $T_r$ by a heuristic rule ({\it Heuristic of task selection}).\label{BI_task_sel}
            \STATE $T_a \leftarrow T_a\!\setminus\!\{i|\exists i,(i,cursor)\in S\}$\label{BI_task_del}
        \ELSE
            \IF {$cursor=E_r$}
                \STATE $tail \leftarrow tail-1$\label{BI_tail_1}
            \ELSE
                \STATE $S \leftarrow$ shift all the pulses of tasks in S one step left.\label{BI_one_left}
                \STATE $S \leftarrow$ BackwardInterleaving($T_a,S,tail,tail\!+\!\min(0,A_{l,S}\!-\!1),p$)\label{BI_call BI_one_left}
                \STATE $tail \leftarrow$ rightmost scheduled slot in $S$\label{BI_tail_2}
            \ENDIF
        \ENDIF
    \ELSE
        \IF {$cursor\neq E_r$}
            \STATE $S \leftarrow$ shift all the pulses of tasks in S to leftmost point.\label{BI_total_left}
            \STATE $S \leftarrow$ BackwardInterleaving($T_a,S,E_l\!+\!tail\!-\!cursor,$ $\min(tail,A_{l,S}\!+\!E_l\!+\!tail\!-\!cursor\!-\!1),p$)\label{BI_call BI_total_left}
            \RETURN $S$\label{BI_force_return}
        \ENDIF
    \ENDIF
\ENDFOR
\RETURN $S$
\end{algorithmic}
\end{algorithm}

\subsubsection{Procedure 2 - BackwardInterleaving}

\begin{figure}[!t]
\centering{
    \subfloat[]{
        \includegraphics[width=.9\columnwidth]{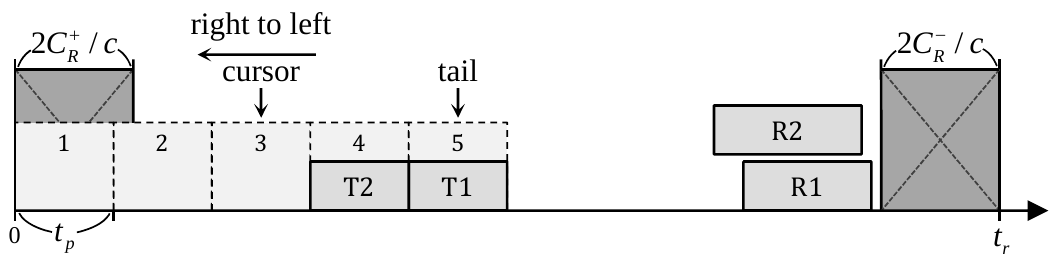}
        \label{fig5a}}\\
    \vspace*{-.1in}
    \subfloat[]{
        \includegraphics[width=.9\columnwidth]{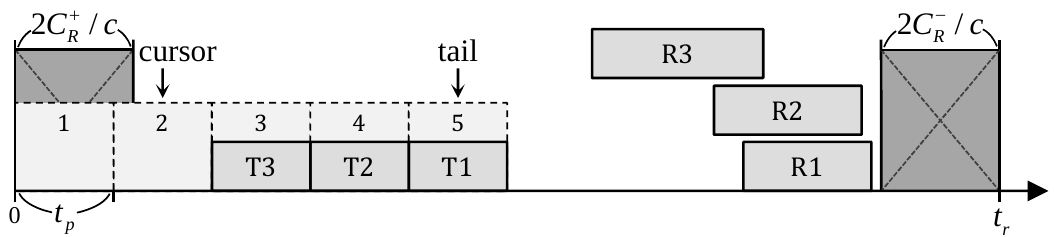}
        \label{fig5b}}\\
    \vspace*{-.1in}
    \subfloat[]{
        \includegraphics[width=.9\columnwidth]{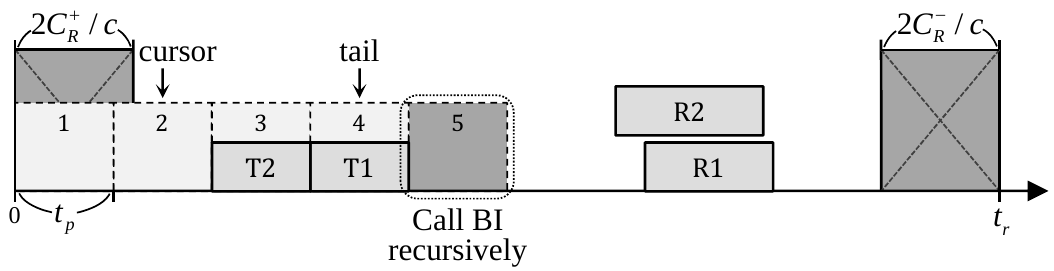}
         \label{fig5c}}\\
    \vspace*{-.1in}
    \subfloat[]{
        \includegraphics[width=.9\columnwidth]{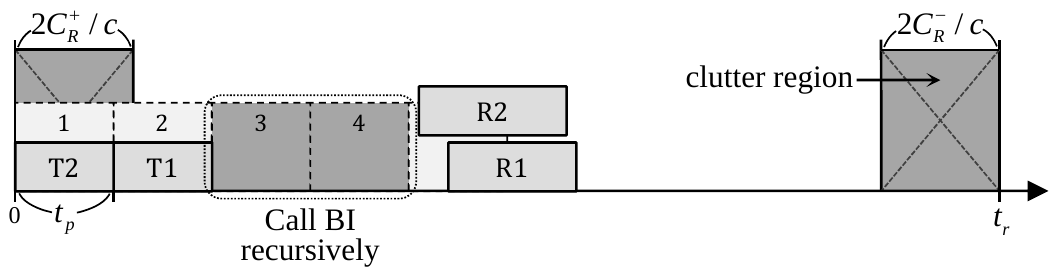}
        \label{fig5d}}}
    \caption{A graphical interpretation of {\alg BackwardInterleaving} algorithm: (a) $cursor$'s movement during iterations of the for-loop; (b) if $T_l\neq\phi$ and $T_r\neq\phi$, a task is selected by a task selection heuristic; (c) if $T_l\neq\phi$ and $T_r=\phi$, the pulses of tasks in $S$ are shifted one slot left; (d) if $T_l=\phi$, the pulses of tasks in $S$ are shifted up to the leftmost slot. T-/R-pulses of tasks are marked with `T' and `R' followed by the task numbers, and the consecutive possible slots of T-pulses are marked with the slot numbers.}
    \label{fig5}
\end{figure}

{\alg BI} is called with the following input variables: the set of available (trackable) tasks $T_a$, a schedule $S$, the leftmost interleaving slot $E_l$, the rightmost interleaving slot $E_r$, and a selected PRF $p$.
$E_l$ and $E_r$ are integer values denoting the slot numbers.
The present scheduling slot and the rightmost scheduled (occupied) slot are denoted by $cursor$ and $tail$, respectively.
$cursor$ and $tail$ are first set to  $E_r$, i.e., the rightmost interleaving slot of a corresponding call of {\alg BI}.
$cursor$ moves from $E_r$ to $E_l$ as the for-loop of {\alg BI} is iterated (see Fig. \ref{fig5a}), and $tail$ is re-assigned right after the rightmost scheduled slot is changed during the call (see Line \ref{BI_tail_1} and \ref{BI_tail_2} of {\alg BI}).

In the for-loop of {\alg BI}, two sets $T_l$, $T_r$ and a value $A_{l,S}$ are calculated:
$T_l$ is the set of tasks, among the tasks in $T_a$, whose leftward availabilities are larger than or equal to the present length of partial schedule, i.e., $(tail-curcor)$;
$T_r$ is the set of tasks, among the tasks in $T_l$, that have rightward availabilities larger than or equal to the present $cursor$;
$A_{l,S}$, the leftward availability of the present schedule $S$, denotes the number of slots where T-pulses of some tasks can be interleaved after the T-pulses of the tasks in $S$: $A_{l,S}=\min_{(i,k)\in S}(k+A_{l,ip(j)})-tail$.
There are three options according to the emptiness of $T_l$ and $T_r$ in the for-loop.
If both $T_l$ and $T_r$ are not empty, there exists at least one task whose T-pulses can be interleaved at $cursor$, then a task is selected by a task selection heuristic (Line \ref{BI_task_sel} and Fig. \ref{fig5b}).
If $T_l$ is not empty but $T_r$ is empty, all the pulses of the tasks in $S$ are shifted only one slot left because there exist some tasks whose T-pulses can be interleaved with the pulses of the tasks in $S$ at smaller $cursor$ (Line \ref{BI_one_left} and Fig. \ref{fig5c}).
If $T_l$ is empty, which means there is no task whose T-pulses and R-pulses can be interleaved with T-pulses of the scheduled tasks, then all the pulses of the tasks in $S$ are shifted up to the leftmost slot (Line \ref{BI_total_left} and Fig. \ref{fig5d}).
After the execution of each statement of the left shifts, another {\alg BI} is called recursively for the slots, on the right side of $S$, that become empty by the left shifts (Line \ref{BI_call BI_one_left} and \ref{BI_call BI_total_left}, and Figs. \ref{fig5c} and \ref{fig5d}).
In case that a {\alg BI} is called after $S$ is shifted up to the leftmost slot, the procedure returns $S$ (Line \ref{BI_force_return}).
Since there are no T-pulses that need to be shifted when $cursor=E_r$, and thus the left shifts and the following recursive {\alg BI}s are meaningless, the two statements of the left shifts are executed only if $cursor$ is not at $E_r$.
%for a finite termination

Since the scheduling in a look is processed backward (leftwards) and a slot at $cursor$ remains empty during scheduling only if no task can be interleaved at the slot, the empty slots to be considered on the right side of $S$ are caused by only the left shifts.
If the one-step left shift in Line \ref{BI_one_left} is executed several times during some iterations of the for-loop, multiple empty slots on the right side of $S$ can be created.
However, the empty slots after $tail$ are no more considered for scheduling, since it is already confirmed that no task exists satisfying the availabilities at those slots by the recursive {\alg BI}s after the earlier left shifts.
Thus, after the one-step left shift of the schedule $S$, the algorithm allows at most a single task to be scheduled at the next right side slot of $S$ by setting $E_l=tail$ and $E_r=tail+\min(0,A_{l,S}-1)$ at the recursive call of {\alg BI} in Line \ref{BI_call BI_one_left}.
On the other hand, if the algorithm shifts $S$ up to the leftmost slot, multiple slots can be cleared at the same time.
%In this case, the slots after $tail$ are not considered for further scheduling since the slots are confirmed to be unavailable by recursive calls of {\alg BI} after the one-step left shifts.
Therefore, {\alg BI} is called recursively in Line \ref{BI_call BI_total_left} with the empty slots from the next right side slot of $S$, i.e., $E_l+tail-cursor$ to the slot of the minimum value between $tail$ and the right most slot where a T-pulse of a new task can be interleaved before R-pulses of tasks in $S$, expressed as $A_{l,S}+E_l+tail-cursor-1$ (see Fig. \ref{fig5d}).

\iffalse
As it is noted at the start of the subsection \ref{sec:alg_edbf}, the finite termination with a feasible solution in the scheduling procedure in a look, i.e. {\alg BI} is a precondition for the algorithm correctness.
The maximum number of the execution of the one-step left shift is at most $N_{\text{intlv}}-1$ since the one-step left shift always followed by a recursively {\alg BI} for the right side slot of the shifted schedule and the slots after $tail$ are no more considered for scheduling.
The execution of the leftmost shift in Line \ref{BI_total_left} is executed at most one time during a call of {\alg BI}, and induces another {\alg BI} for the slots before $tail$,
\fi

\subsubsection{PRF and Task Selection Heuristics}\label{sec:prf_task_rule}

A PRF of an interleaved look in the while-loop of {\alg HIED} and a task scheduled in the for-loop of {\alg BI} are selected by priority index based heuristic rules.
In this paper, greedy (G), reverse greedy (RG) and random (R$_{\text{PRF}}$) selections are used for the PRF selection heuristics in Line \ref{HIED_PRF_sel} of {\alg HIED}.
In every iteration of the while-loop in {\alg HIED}, the greedy heuristic selects the PRF with which the maximum number of unscheduled targets that can be tracked.
The greedy heuristic is a well-known $\log n$-approximation algorithm for the set-cover problem and it is also proved as a $\log n$-approximation algorithm for the general uncapacitated facility location problem \cite{Hoc82}.
The reverse greedy and the random rules are used for comparison: the reverse greedy selects a PRF with the minimum number of trackable targets, and the random rules selects a PRF randomly.

For the task selection heuristics in Line \ref{BI_task_sel} of {\alg BI}, 6 priority index based rules are used:
\begin{itemize}
\item shortest ambiguous range first (SAR): select a task of a target with the smallest $R_a$.
\item longest ambiguous range first (LAR): select a task of a target with the largest $R_a$.
\item random (R$_{\text{task}}$): select a task randomly.
\item smallest number of available PRFs first (SAP): select a task with the smallest number of PRFs that the task can be performed with.
\item smallest leftward availability first (SLA): select a task with the smallest sum of its leftward availabilities for all PRFs.
\item smallest rightward availability first (SRA): select a task with the smallest sum of its rightward availabilities for all PRFs.
\end{itemize}

Recall that a PRI is the interval between the start times of two successive T-pulses, and thus the ambiguous range of a target can be used to roughly estimate the lengths of intervening idle times before and after the R-pulse in a PRI.
If the ambiguous range is short, preferable in SAR rule, the former intervening idle time is short whereas the latter becomes long.
Then the pulses of the task(target) is more suitable to be interleaved within T-/R-pulses of other tasks rather than to be interleaved with the pulses of other tasks.
On the contrary, LAR rule selects a task with long ambiguous range first, which might have a long intervening idle time before the R-pulse and thus be suitable to be interleaved with other tasks.

The above ambiguous range based heuristic rules take into account the intervening idle times between pulses, but they do not reflect exactly the options of interleaving for a task.
The rules using PRF availability information such as SAP, SLA, and SRA are based on the numbers of the options for a task.
In a task selection call at a certain slot with some PRF, a task that can be tracked with many other PRFs or can be scheduled at many other slots still has chances to be interleaved with other tasks in other looks and slots, but a task having smaller options of interleaving is relatively difficult to be interleaved with other tasks in other conditions and might increase the total number of looks for interleaving unless it is scheduled at the slot.
Therefore, SAP, SLA, and SRA select a task with the smallest options of interleaving, i.e. smallest sums of availabilities.

\subsection{Heuristic Interleaving Algorithm for SDBF}\label{sec:alg_sdbf}

\begin{figure}[t]
\centerline{
    \includegraphics[width=0.75\columnwidth]{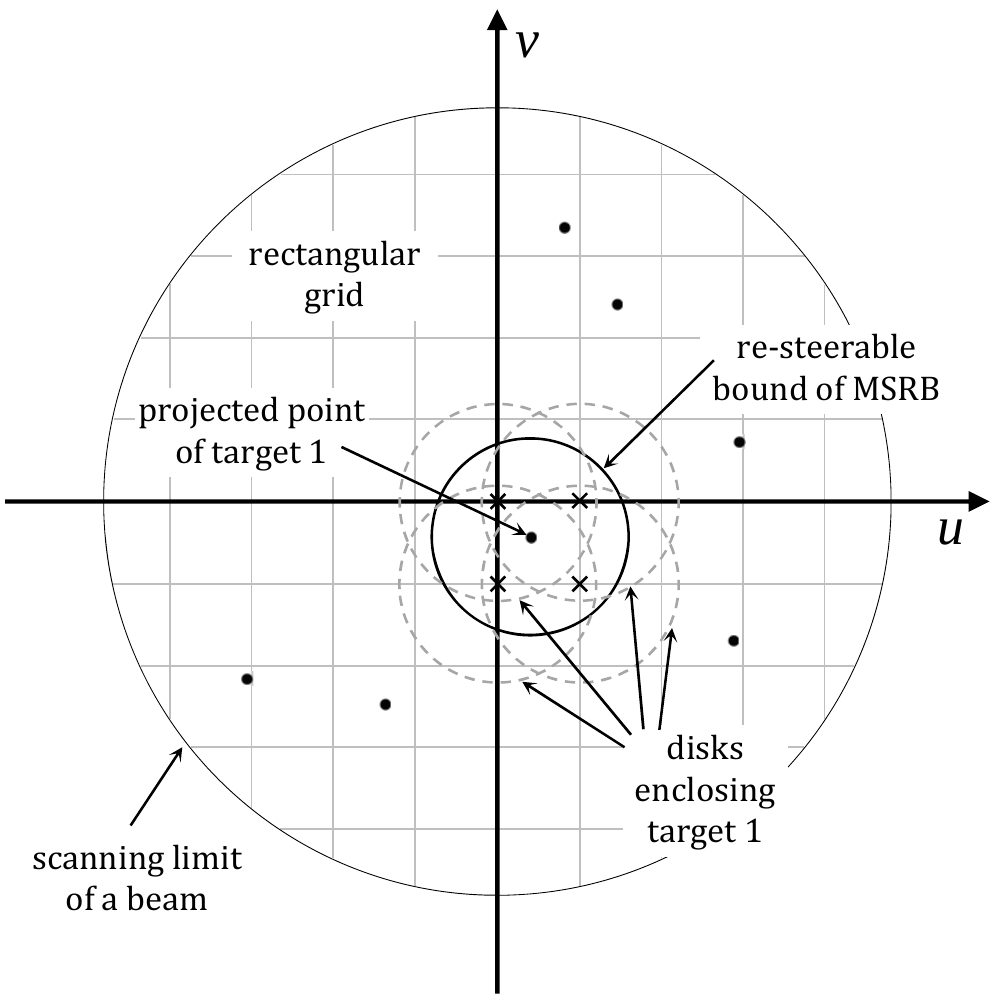}}
    %\vspace*{.01in}
    \caption{The disks (dotted gray) on a rectangular grid enclosing a projected point of a target trackable with a PRF: the grid points within the disk ($\equiv$ the re-steerable bound of MSRB) radius from the projected target point are utilized as the disk centers (marked as X). $u$ and $v$ are the horizontal and vertical domains of the plane.}
    \label{fig6}
    %\vspace*{-.1in}
\end{figure}

The heuristic interleaving algorithm for the SDBF case is characterized by preprocessing of looks with disks and disk selection of an interleaved look, compared with the algorithm for the EDBF case.
As it is noted in section \ref{sec:pf_sdbf}, although an exact unique disk search process gives a minimum cardinality collection of disks for identifying assential looks of the interleaving problem,
this paper utilizes a rectangular grid on the normalized scanning plane, which discretizes the solution space of unique disk search, as an approximated approach for identifying the looks;
the grid points within the disk radius from any projected target point on the normalized scanning plane are used as centers of disks for the heuristic pulse interleaving (see Fig. \ref{fig6}).

\iffalse
As it is noted in section \ref{sec:bg_msrb}, a re-steerable area of MSRB of a look is regarded as a disk enclosing the projected points on the normalized scanning plane of some targets that can be trackable through the look.
If radar beams is steerable in a continuous manner, any point in the normalized scanning plane can be the center of a disk.
Then, in principle, there can be infinite number of disks and corresponding looks to be considered as a problem space of the interleaving problem.
Since multiple disks
\fi

After the preprocessing for all PRFs, the algorithm selects some disk of a look according to a heuristic rule and try to interleave and schedule the tasks, of which target is located in the selected disk, into the look.
This process is repeated until all tasks are assigned.
For the pulse scheduling procedure in a look with a selected disk, {\alg BI} is used as in the EDBF case and thus the correctness of the algorithm in the SDBF case can be handled in the same way.

\subsubsection{Procedure 3 - HeuristicInterleavingSubarrayDBF}

Procedure \ref{alg3} ({\alg  HeuristicInterleavingSubarrayDBF} or {\alg HISD}) is the pseudocode for the preprocessing of looks with disks and the disk selection.
First, disks centered on grid points that enclose at least one projected target point are identified in the cascade of for-loops (Line \ref{HISD_for_1_start}-\ref{HISD_for_1_end}).
For each task $i$ trackable with a PRF $p$, the set $D_i$ of disks that enclose the projected target point $q_i$ with given disk radius $r$, the center $g_d$ of each disk $d$, and PRF $p_d=p$ of the look corresponding to $d$ is calculated.
%If the projected target point $q_i$ of a task $i$ is within the disk radius $r$ from the center $g_d$ of a disk $d$ and task $i$ is trackable with PRF $p_d$ of the look corresponding to disk $d$, the projected target point $q_i$ of task $i$ is enclosed by disk $d$ centered on $g_d$ in the normalized scanning plane.
Then, for each disk $d$ in $D_i$, the task index $i$ is added to a task list $T_d$, initialized as the set of tasks whose projected target points are enclosed by $d$ through the entire cascade of for-loops.
By this process, all nonempty disks on the grid for each PRF are identified without unnecessary computation of empty disks (for the computational complexity, see \ref{sec:complex_sdbf}).
%To identify all nonempty disks for each PRF, {\alg HISD} inserts projected target points of tasks in the normalized scanning plane one by one, and adds the task index to a task list $T_d$ for each disk in $D_i$ (Line \ref{HISD_T_d}).

\begin{algorithm} [t]
    \caption{HeuristicInterleavingSubarrayDBF}
    \label{alg3}
\begin{algorithmic}[1]
%\STATE Load $A_{v,in}$, $A_{l,in}$, and $A_{r,in}$ for all $(i,n)$ pairs.\label{HISD_load_av}
\FOR {$p \leftarrow 1$ to $N_{\text{PRF}}$}\label{HISD_for_1_start}
    \STATE $T_{v,p} \leftarrow \{i|\forall i,A_{v,ip(j)}=1\}$
    %\FOR {$d\in\{d|\forall d,p_d=p\}$}
        %\STATE $T_d=\{i\in T_{v,p}|\|g_d-q_i\|<r\}$\label{HISD_T_d}
    %\ENDFOR
    \FOR {$i\in T_{v,p}$}
        \STATE  $D_i \leftarrow \{d|\forall d,p_d=p,\|g_d-q_i\|\leq r\}$\label{HISD_D_i}
        \FOR {$d\in D_i$}
            \STATE  $T_d\leftarrow T_d\cup\{i\}$ \label{HISD_T_d}
        \ENDFOR
    \ENDFOR
\ENDFOR\label{HISD_for_1_end}
\STATE $H \leftarrow \phi$\label{HISD_init_start}
\STATE $T_0 \leftarrow$ the set of all tasks $\equiv \{i|\forall i\}$
\STATE $j \leftarrow 0$\label{HISD_init_end}
\WHILE {$T_0\neq\phi$}
    \STATE $j \leftarrow j+1$
    \STATE $d \leftarrow$ select a disk by a heuristic rule ({\it Heuristic of disk selection}).\label{HISD_disk_sel}
%    \STATE $p \leftarrow p_d$
%    \STATE $T_a \leftarrow \{i|\forall i\in T_d\}$
    \STATE $S \leftarrow \phi$
    \STATE $S \leftarrow$ BackwardInterleaving($T_d,S,1,N_{\text{intlv}},p_d$)\label{HISD_call_BI}
    \STATE $T_0 \leftarrow T_0\!\setminus\!\{i|\forall (i,k)\in S\}$\label{HISD_task_del_0}
    \STATE $\forall d, \: T_d \leftarrow T_d\!\setminus\!\{i|\forall (i,k)\in S\}$\label{HISD_task_del_d}
    \STATE $H \leftarrow H\cup\{(i,j,k)|\forall (i,k)\in S\}$\label{HISD_sch}
\ENDWHILE
\RETURN $H$
\end{algorithmic}
\end{algorithm}

The disks on the grid identified in {\alg HISD} are inherently duplicative and redundant: since a single projected target point can be enclosed by multiple disks, multiple disks can have the same task list $T_d$ and some disks can have task lists that include task lists of other disks.
These duplicative and redundant disks, i.e., duplicative and redundant looks increase unnecessary calculation for solving the integer program, thus should be arranged before the integer programming so that no disks are duplicative and no disks are subset of others in terms of their task lists.
However, this process is not necessary for the heuristic interleaving algorithm since a priority index based disk selection heuristic selects a disk among the `best' disks and scheduled tasks are eliminated from the task lists of all the disks after the pulse scheduling (Line \ref{HISD_task_del_d}).
The retention of the duplicative and redundant disks increases the computational load in the while-loop of {\alg HISD}, but a larger computation is required to discriminate supersets among the disks for eliminating duplicates and subsets.

After the disks are created, the interleaving result $H$, the set of all tasks $T_0$, and look index $j$ are initialized in Line \ref{HISD_init_start}-\ref{HISD_init_end}.
At every iteration of the while-loop in {\alg HISD}, a disk, whose PRF and look indices are $p_d$ and $j$, is selected by a disk selection heuristic in Line \ref{HISD_disk_sel}, and then {\alg BI} is called for the pulse scheduling in Line \ref{HISD_call_BI}, as is the case in EDBF.
The scheduled tasks in $S$ are eliminated from $T_0$ and all $T_d$s after the pulse scheduling, and triplets $(i,j,k)$ of the scheduled tasks are inserted in $H$ (Line \ref{HISD_task_del_0}-\ref{HISD_sch}).
Repeating this procedure in the while-loop, if $T_0$ becomes empty, then {\alg HISD} returns $H$ and terminates.

\subsubsection{Disk Selection Heuristics}

A disk selection heuristic in {\alg HISD} consists of two priority indices: a main index and a sub-index for tie-breaking.
Since the main indices used in this paper are based on the cardinalities of disks and  there can exist many disks of the same cardinality, including duplicates, at least randomized tie-breaking is needed to avoid initial biases.
As main indices, greedy disk (GD), reverse greedy disk (RGD), and weighted greedy disk (WGD) selections are used in this paper.
Similarly to the greedy and reverse greedy heuristics in the EDBF case, GD selects a disk of the maximum cardinality, i.e., the maximum number of enclosed tasks, whereas RGD selects a disk of the minimum cardinality.
In WGD, a disk of the maximum weighted cardinality is selected, where the weight is the sum of the reciprocals of the number of available disks for targets enclosed by the disk.
By this weight, the disk that encloses targets having minimal options to be interleaved in other looks(disks) is preferentially selected.
The sub-indices used for tie-breaking are random (R$_{\text{disk}}$) and smallest dwell time first (SD).
R$_{\text{disk}}$ is random selection of disks among the disks with the same value of the main index, and SD selects a disk whose look has smallest dwell time among the disks: the latter is for reducing the total tracking time as far as possible.

\section{Algorithm Complexity Analysis}\label{sec:alg_complex}

The asymptotic time complexities of the interleaving algorithms are analyzed in this section.
The algorithms can be implemented in several ways with some data structures for the PRF/disk and task selection heuristics.
In PRF selection, R$_{\text{PRF}}$ utilizes randomly generated priority indices, but G and RG use the cardinality of the set of unscheduled tasks trackable with a PRF $p$, i.e. $|T_p|$ as their priority index.
Since $|T_p|$ is an integer and increases/decreases with insertion/deletion of tasks, it will be shown in the following subsection that a PRF selection process using G or RG, as in random number generation, takes $O(1)$ time with a particular sorted list.
By the same token, disk selection heuristics composed of GD or RGD as the main index and R$_{\text{disk}}$ as the sub-index require constant time for each execution (see \ref{sec:complex_sdbf}).
But, if the disk heuristic uses a continuous priority index (WGD) or a sub-index that requires comparison operations (SD), a search tree is needed for an efficient computation (see \ref{sec:complex_sdbf}).
In {\alg BI}, a task among the tasks in $T_r$ is selected so that pulses of the task are interleaved, and thus a task selection heuristic based on priority index must be valid in different sets of tasks, i.e. $T_l$s and $T_r$s.
Three options for the task selection are analyzed in this paper : brute force search through an unsorted task list (\ref{sec:complex_bfs}), selection from $N_{\text{intlv}}(N_{\text{intlv}}-1)$ sorted task lists of all pairs of $(T_l,T_r)$ for each PRF (\ref{sec:complex_stl}), and multi-level search by 3-dimensional orthogonal range trees (\ref{sec:complex_mls}).

\subsection{Complexity of Interleaving Algorithm for EDBF with Brute Force Search}\label{sec:complex_bfs}

First, the time complexity of the pulse interleaving algorithm for EDBF using the brute force search in the task selection is analyzed.
In {\alg HIED}, to load and calculate availabilities of all task-PRF pairs, $O(N_tN_{\text{PRF}})$ time is spent.
The initiation of $T_0$ in Line \ref{HIED_T_0} requires $O(N_t)$ time computation, and the while-loop runs at most $N_t$ times in the worst cases.
For PRF selection heuristics, a sorted list of PRFs whose length is $N_{\text{PRF}}$ is used.
Each of the PRFs in the list has a pointer to a value of $|T_p|$ and vice versa, and thus multiple PRFs of equal $|T_p|$s can have the same pointers.
Each value of $|T_p|$ appended to the list is mutually connected with the next larger value and the next smaller value; the maximum and the minimum value of $|T_p|$ have self-pointing addresses.
Initially, all PRFs point to 0, and inserting tasks one by one, the pointer of every PRF available for the inserted task moves to the value of $|T_p|$ increased by 1.
If the value does not exist, then it is created and connected to the PRF's former value of $|T_p|$ and the next larger value.
When a value of $|T_p|$ has no pointer to any PRFs, the value is deleted and the next larger and smaller values of it are mutually connected.
The computation time for this preprocessing is proportional to $\sum_{p=1}^{N_{\text{PRF}}}|K_p|$, hereafter denoted by $Q_p$, where $K_p$ is the set of all tasks trackable with a PRF $p$.
The PRF heuristic selects one of the PRFs with the largest value of the list (G) or with the smallest value (RG) and thus it takes $O(1)$ time.
During the preprocessing, the task list $T_p$ of each PRF is also constructed within the same complexity $Q_p$ and every task is mutually connected with its available PRFs for task deletion after the pulse scheduling in a look.
By this task deletion from $T_p$, no extra operation is needed for Line \ref{HIED_T_p}.

\begin{Prop}\label{prop1}
The sorted list of PRFs with the cardinalities of their trackable tasks for constant time PRF selection can be constructed in $O(\sum_{p=1}^{N_{\text{PRF}}}|K_p|)=O(Q_p)$ time, and a corresponding task list can be constructed in $O(Q_p)$ time.
\end{Prop}

At every call of {\alg BI} from Line \ref{HIED_call_BI} in {\alg HIED}, the for-loop of {\alg BI} runs exactly $N_{\text{intlv}}$ times unless the forced return in Line \ref{BI_force_return} of {\alg BI} is executed.
Since a recursive call of {\alg BI} in Line \ref{BI_call BI_one_left} is with at most a single interleaving slot, the recursive {\alg BI} ends in a single for-loop and the corresponding slot is occupied or abandoned in the further process.
Therefore, this recursion additionally iterates the for-loop at most $N_{\text{intlv}}$ times during a call of {\alg BI} from {\alg HIED}.
On the other hand, the recursive call of {\alg BI} in Line \ref{BI_call BI_total_left} does not increase the total iteration of the for-loop since the call of {\alg BI} in Line \ref{BI_call BI_total_left} is with unvisited remaining interleaving slots after the total left shift and the algorithm returns right after Line \ref{BI_call BI_total_left}.

\begin{Prop}\label{prop2}
The for-loop of {\alg BI} iterates at most $2N_{\text{intlv}}$ times during a call of {\alg BI} from {\alg HIED}.
\end{Prop}

As a task $i$ is selected in Line \ref{BI_task_sel} of {\alg BI}, the leftward availability of the present schedule, i.e. $A_{l,S}$ is changed to the minimum between the previous $A_{l,S}$ and $cursor +A_{l,ip(j)}-tail$:
the latter is the number of slots between the T-pulses of the present schedule $S$ and the R-pulse of $i$.
Therefore, $A_{l,S}$ in Line \ref{BI_A_lS}, which is initially $N_{\text{intlv}}$ by definition, can be updated in $O(1)$ time after every task selection.
The left shifts in Line \ref{BI_one_left} and \ref{BI_total_left} can be done in $O(1)$ by an appropriate list of pointers for interleaved slots and corresponding tasks in $S$:
since $S$ is composed of at most two lists of tasks whose T-pulses are consecutive by the left shifts, 4 pointers for both ends of the two task lists are enough to represent and manipulate interleaved slots for the shifts.

$T_l$ and $T_r$ are the sets of tasks used for emptiness check in if-statements of {\alg BI} and for heuristic task selection.
Thus, a list of tasks that satisfy the conditions of $T_l$ and $T_r$ is required to be identified for the task selection.
In brute force search, in every iteration of the for-loop of {\alg BI}, all the tasks in $T_a$ are checked whether they satisfy the conditions of $T_l$ and $T_r$ in order, and then a task with the maximum priority index among the tasks in $T_r$ is searched and selected.
Therefore, the operations for brute force search require $O(|T_p|)$ time in the worst cases.

\begin{Prop}\label{prop3}
A call of {\alg BI} from {\alg HIED} totally requires $O(N_{\text{intlv}}|T_p|_{\max})$ time by the brute force search even in the worst cases, where $|T_p|_{\max}\equiv \max_p|T_p|$.
\begin{proof}
Since it is already shown that other operations in the for-loop of {\alg BI} except the brute force search takes $O(1)$ time, the time complexity of a single execution of the for-loop is $O(|T_p|)$ in summary.
By Proposition \ref{prop2}, the number of iterations of the for-loop is $O(N_{\text{intlv}})$, and thus a call of {\alg BI} from {\alg HIED} takes $O(N_{\text{intlv}}|T_p|_{\max})$ time by the brute force search even in the worst cases
\end{proof}
\end{Prop}

%, a call of {\alg BI} from {\alg HIED} totally requires $O(N_{\text{intlv}}|T_p|_{\max})$ in the worst cases

After the pulse scheduling in a look, the scheduled tasks are deleted from $T_p$s of all PRFs and $T_0$, and merged into $H$: the time for merge of each task is trivially $O(1)$.
Since a task is mutually connected to all available PRFs, the deletion takes $O(|P_i|+1)$ time, where $P_i$ denotes the set of available PRFs of task $i$.
The deletion of the task reduces $|T_p|$ of the corresponding PRFs, and therefore the pointers of the PRFs are moved to the values decreased by 1 in the sorted PRF list; all the related operations are similar to the preprocessing.
\begin{Lem}\label{lem1}
The time complexity of the pulse interleaving algorithm for EDBF using the brute force search is $O(N_t(N_{\text{PRF}}+N_{\text{intlv}}|K_p|_{\max}))$, where $|K_p|_{\max}\equiv \max_p|K_p|$.
\begin{proof}
The sum of the cardinalities of all $K_p$s and the sum of the cardinalities of all $P_i$s are the same: $\sum_{p=1}^{N_{\text{PRF}}}|K_p|=\sum_{i=1}^{N_t}|P_i|=Q_p$.
The deletion and the merge occur, respectively, only one time per task, and thus $O(Q_p+N_t)=O(Q_p)$ time is needed for the deletion and the merge during the entire algorithm execution.
The calculations for the availabilities of all task-PRF pairs takes $O(N_tN_{\text{PRF}})$ time, and the while-loop of {\alg HIED} runs at most $N_t$ times.
Therefore, by Proposition \ref{prop1} and \ref{prop3}, the time complexity of the algorithm with the brute force search is $O(N_tN_{\text{PRF}}+N_t\cdot N_{\text{intlv}}|K_p|_{\max}+Q_p)=O(N_t(N_{\text{PRF}}+N_{\text{intlv}}|K_p|_{\max}))$ since $T_p\subseteq K_p$ and $Q_p\leq N_tN_{\text{PRF}}$.
\end{proof}
\end{Lem}

\subsection{Sorted Task Lists}\label{sec:complex_stl}

The second option of the data structure for the task selection heuristics is to prepare a sorted task list for each pair of  $(T_l,T_r)$.
By the definition in Line \ref{BI_T_l} and Line \ref{BI_T_r} of {\alg BI}, $T_l$ and $T_r$ are the sets of tasks whose $A_{l,ip(j)}$ and $A_{r,ip(j)}$ are larger than certain positive integers.
Since $0\leq A_{l,ip(j)}\leq N_{\text{intlv}}$ and $1\leq A_{r,ip(j)}\leq N_{\text{intlv}}$ for a task $i\in K_p$, there are $N_{\text{intlv}}(N_{\text{intlv}}-1)$ possible pairs of $(T_l,T_r)$.
In the preprocessing for the second option, the tasks satisfying the conditions of each $(T_l,T_r)$ pair are sorted according to a priority index of a task selection heuristic: it is done in $O(|K_p|\log|K_p|)$ time.
For all PRFs and all $(T_l,T_r)$ pairs, the time complexity of the preprocessing is $O(\sum_{p=1}^{N_{\text{PRF}}}(|K_p|\log|K_p|)N_{\text{intlv}}^2)$ or less tightly $O(Q_pN_{\text{intlv}}^2\log|K_p|_{\max})$.

The complexities of most operations in the algorithm are the same as the brute force search except the identification of $T_l$ and $T_r$ (Line \ref{BI_T_l} and \ref{BI_T_r} of {\alg BI}) and task selection and deletion (Line \ref{BI_task_sel} and \ref{BI_task_del} of {\alg BI} and Line \ref{HIED_task_del} of {\alg HIED}).

\begin{Prop}\label{prop4}
The time complexity of a call of {\alg BI} from {\alg HIED} is $O(N_{\text{intlv}}^2N_s)$ using multiple sorted task lists, where $N_s$ denotes the number of scheduled tasks in the call.
\begin{proof}
Since $T_l$ and $T_r$ are identified in advance, no extra operation is needed for Line \ref{BI_T_l}, and \ref{BI_T_r} of {\alg BI} and task selection is also done in $O(1)$ time from the sorted list.
However, a selected task has to be deleted from all $N_{\text{intlv}}(N_{\text{intlv}}-1)$ sorted lists of the selected look's PRF.
The deletion can be processed in $O(N_{\text{intlv}}^2)$ time with task's pointers to the task lists, and thus the deletion of all scheduled tasks takes $O(N_{\text{intlv}}^2N_s)$ during a call of {\alg BI} from {\alg HIED}.
Since other operations in {\alg BI} take $O(1)$ time for each iteration of the for-loop, and by Proposition \ref{prop2}, the time complexity of a call of {\alg BI} from {\alg HIED} is $O(N_{\text{intlv}}^2N_s)$.
\end{proof}
\end{Prop}

\begin{Lem}\label{lem2}
The time complexity of the pulse interleaving algorithm for EDBF using multiple sorted task lists is $O(N_tN_{\text{PRF}}+Q_pN_{\text{intlv}}^2\log|K_p|_{\max})$.
\begin{proof}
For each scheduled task, the deletion from the task lists of other PRFs is also required after pulse scheduling: the time complexity is $O(|P_i|N_{\text{intlv}}^2)$ per task.
Therefore, the task deletion through the algorithm requires $O(\sum_{i=1}^{N_t}|P_i|N_{\text{intlv}}^2)=O(Q_pN_{\text{intlv}}^2)$ time.
Since a task is scheduled only one time in the algorithm, the time spent by all the calls of {\alg BI} from {\alg HIED} is $O(N_tN_{\text{intlv}}^2)$.
The calculations for the availabilities of all task-PRF pairs takes $O(N_tN_{\text{PRF}})$ time, and the time complexity of the preprocessing for the sorted task lists is $O(Q_pN_{\text{intlv}}^2\log|K_p|_{\max})$.
In summary, the time complexity of the algorithm using multiple sorted task lists is $O(N_tN_{\text{PRF}}+Q_pN_{\text{intlv}}^2\log|K_p|_{\max}+N_tN_{\text{intlv}}^2+Q_pN_{\text{intlv}}^2)=O(N_tN_{\text{PRF}}+Q_pN_{\text{intlv}}^2\log|K_p|_{\max})$ since $Q_p\geq N_t$.
\end{proof}
\end{Lem}

\subsection{A Multi-Level Search by Orthogonal Range Trees}\label{sec:complex_mls}

The sorted task lists for task selection can be stored more efficiently by orthogonal range trees.
Since a task with the maximum priority index is selected among the tasks whose $A_{l,ip(j)}$ and $A_{r,ip(j)}$ are larger than certain positive integers, a data structure is utilizable that can efficiently answer orthogonal rectangular range queries in the $A_{l,ip(j)}\!-\!A_{r,ip(j)}$ plane and efficiently report a task with the maximum priority index from the query result.
A 3-dimensional orthogonal range tree is appropriate for the requirements:
the first level of this range tree is a balanced binary search tree of $A_{l,ip(j)}$;
for each node in the first level tree, an auxiliary search tree of $A_{r,ip(j)}$ is built as the second level tree;
the last level of the range tree needs not to be a search tree but just to be a sorted list of tasks according to a priority index.

\begin{Prop}\label{prop5}
The orthogonal range tree can answer the task selection/deletion query in $O(\log^2N_{\text{intlv}})$ time and can be built in $O(|K_p|(\log|K_p|+\log^2N_{\text{intlv}}))$ time for each PRF.
\begin{proof}
The depths of the first and second level trees are at most $\log N_{\text{intlv}}$ since $A_{l,ip(j)}$ and $A_{r,ip(j)}$ are positive integers no larger than $N_{\text{intlv}}$.
Then, the preprocessing for the range trees is as follows.
First, for each PRF, sort tasks by a priority index of a task selection heuristic in $O(|K_p|\log|K_p|)$ time.
Then, insert the tasks to the range tree in the sorted order in $O(|K_p|\log^2N_{\text{intlv}})$ time.
As a result, the tasks at the third level in each node of the second level search tree are naturally sorted.

Using this orthogonal range tree, the identification of $T_l$ and $T_r$ in {\alg BI} can be done in $O(\log N_{\text{intlv}})$ and $O(\log^2N_{\text{intlv}})$, respectively.
To acquire the best task according to the priority index, all the best tasks from task lists of $\log^2N_{\text{intlv}}$ nodes of the second level trees, reported by the range query of $T_r$, are compared.
Obviously, this comparison is done in $O(\log^2N_{\text{intlv}})$ time.
The deletion of the selected task is processed in $O(\log^2N_{\text{intlv}})$ time since the task can be contained in $\log^2N_{\text{intlv}}$ nodes through the orthogonal range tree.
Therefore, the task selection followed by the deletion is processed in $O(\log^2N_{\text{intlv}})$ time.
\end{proof}
\end{Prop}

\begin{Prop}\label{prop6}
The time complexity of a call of {\alg BI} from {\alg HIED} is $O(N_{\text{intlv}}\log^2N_{\text{intlv}})$.
\begin{proof}
The identification of $T_r$ can occur in every iteration of the for-loop of {\alg BI}, which takes $O(N_{\text{intlv}}\log^2N_{\text{intlv}})$ in total, whereas the task selection and deletion are processed $N_s(\leq N_{\text{intlv}})$ times.
Other operations in {\alg BI} takes O(1) time for each iteration of the for-loop, and thus the proposition is concluded.
\end{proof}
\end{Prop}

\begin{Lem}\label{lem3}
The time complexity of the pulse interleaving algorithm for EDBF using orthogonal range trees is $O(N_tN_{\text{PRF}}+Q_p(\log|K_p|_{\max}+\log^2N_{\text{intlv}})+N_tN_{\text{intlv}}\log^2N_{\text{intlv}})$.
\begin{proof}
The scheduled tasks are deleted from the orthogonal range trees of other PRFs after the pulse scheduling and the time complexity of this deletion is $O(|P_i|\log^2N_{\text{intlv}})$ per task.
This task deletion through the algorithm takes $O(\sum_{i=1}^{N_t}|P_i|\log^2N_{\text{intlv}})=O(Q_p\log^2N_{\text{intlv}})$ time.
Since the while-loop of {\alg HIED} runs at most $N_t$ times, the time spent by all the calls of {\alg BI} from {\alg HIED} is $O(N_tN_{\text{intlv}}\log^2N_{\text{intlv}})$ by Proposition \ref{prop6}.
The calculations for the availabilities of all task-PRF pairs takes $O(N_tN_{\text{PRF}})$ time, and the time complexity of the preprocessing for the orthogonal range trees for all PRFs is $O(Q_p(\log|K_p|_{\max}+\log^2N_{\text{intlv}}))$ by Proposition \ref{prop5}.
In summary, the time complexity of the algorithm using orthogonal range trees is $O(N_tN_{\text{PRF}}+Q_p(\log|K_p|_{\max}+\log^2N_{\text{intlv}})+N_tN_{\text{intlv}}\log^2N_{\text{intlv}})$.
\end{proof}
\end{Lem}

With a small constant $N_{\text{intlv}}$, as a reasonable assumption with a high duty ratio (see \ref{sec:bg_msrb}), $N_tN_{\text{intlv}}\leq Q_p\log|K_p|_{\max}$.
Thus, Lemma \ref{lem3} shows that, In comparison with the complexity of the algorithm using multiple sorted lists, the algorithm runs faster by orthogonal range trees in terms of the asymptotic complexity.

The time complexities shown in Lemma \ref{lem2} and \ref{lem3} can be simplified with some constant parameters and a scalable number of targets.

\begin{Thm}\label{thm1}
The time complexity of the pulse interleaving algorithm for EDBF using multiple sorted task lists or orthogonal range trees is $O(N_t\log N_t)$ if $N_{\text{PRF}}$ and $N_{\text{intlv}}$ are constant.
\begin{proof}
Since $|K_p|_{\max}\leq N_t$, if $N_{\text{PRF}}$ and $N_{\text{intlv}}$ are constant, then $O(Q_p)\leq O(N_tN_{\text{PRF}})=O(N_t)$ and the complexities in Lemma \ref{lem2} and \ref{lem3} are reduced to $O(N_t\log N_t)$.
\end{proof}
\end{Thm}

\subsection{Complexity of Interleaving Algorithm for SDBF}\label{sec:complex_sdbf}

The heuristic interleaving algorithm for SDBF produces a number of disks on a grid, which increase the computation time and require extra data structures, in the preprocessing of looks.
The disks are created by inserting projected target points of tasks on the normalized scanning plane for each PRF.
The grid points of a rectangular grid with a spacing $\epsilon$ can be the centers of the disks if some projected target point is located within the disk radius $r$ from the grid points (see Fig. \ref{fig6}).

Thus, the preprocessing of the disks is as follows.
First, for each PRF, inserting trackable tasks one by one, a region of grid points, that can be the centers of the disks, is calculated;
The region consists of $\lfloor\frac{2r}{\epsilon}\rfloor$ row intervals of abscissa or column intervals of ordinate.
Then, two data structures are built for the disks during the insertion of tasks.
The first is a search tree of the disks, for each PRF, sorted lexicographically by the coordinates of the disks' centers.
This search tree is to identify whether the disks already exist whose centers are in the region of disk centers from a new task insertion.
The second is a sorted list of the disks for all PRFs according to the cardinality of disk: the list has the same structure as the one used for PRF selection heuristics.

Let $N_{d,p}$ denote the number of disks for PRF $p$, $N_d(\equiv\sum_{p=1}^{N_{\text{PRF}}}N_{d,p})$ be the number of disks of all PRFs, $K_d$ be the set of all tasks trackable in a disk $d$, $P_i'$ be the set of available disks of task $i$, and $Q_d$ denote the sum of the cardinalities of all $K_d$s, i.e. $Q=\sum_{d=1}^{N_d}|K_d|=\sum_{i=1}^{N_t}|P_i'|$.

\begin{Prop}\label{prop7}
The search trees and the sorted list for disks on a rectangular grid are constructed in $O(N_d\log N_{d,p}|_{\max}+Q_d)$ time.
\begin{proof}
The search trees are constructed during the preprocessing of the disks on the grid to identify existing disks in the trees by previous task insertions.
Since the region of the centers of disks enclosing an inserted target point of a task are calculated in $O(r/\epsilon)$ time for each task insertion, where $r/\epsilon\leq P_i'$, and each identified disk $d$ encloses $|K_d|$ projected target points, at least $O(Q_d)$ time is required for identifying all the disks in the preprocessing.
The search tree for each PRF is obviously constructed in $O(N_{d,p}\log N_{d,p})$ time, and thus the processing of all of those trees requires $O(\sum_{p=1}^{N_{\text{PRF}}}(N_{d,p}\log N_{d,p}))$ time or less tightly $O(N_d\log N_{d,p}|_{\max})$ for all PRFs.
By the same way of constructing the sorted list of PRFs in \ref{sec:complex_bfs}, the sorted list of the disks for all PRFs with their cardinality is constructed in $O(Q_d)$ time.
Thus, the two types of the structures are constructed in $O(N_d\log N_{d,p}|_{\max}+Q_d)$.
\end{proof}
\end{Prop}

For WGD, the weight of a disk, i.e. the sum of the reciprocal of the number of available disks for the tasks enclosed by the disk is also calculated.
Since the number of available disks of a task is trivially known through the task insertion, no extra complexity increases.
However, the disks are must be sorted by the continuous weighted cardinality in $O(N_d\log N_d)$ time.
Any sub-index of a disk selection heuristic except random selection also needs sorting of disks, and thus $O(N_d\log N_d)$ computation is required.

The number $N_{d,p}$ of disks for PRF $p$ is proportional to the disk area, the density of grid points, and the number of the task insertions, and thus $N_{d,p}=O(\frac{r^2}{\epsilon^2}|K_p|)$.
Therefore, the following proposition is derived.

\begin{Prop}\label{prop8}
The time complexity of the preprocessing for the disks is $O(\frac{r^2}{\epsilon^2}Q_p\log\frac{r^2}{\epsilon^2}|K_p|_{\max}+Q_d)$ if GD or RGD without any sub-index is used for a disk selection heuristic, and it is $O(\frac{r^2}{\epsilon^2}Q_p\log\frac{r^2}{\epsilon^2}Q_p+Q_d)$ if WGD or any sub-index is used for a disk selection heuristic.
\begin{proof}
If GD or RGD without any sub-index is used for a disk selection heuristic, by Proposition \ref{prop7}, the time complexity of the preprocessing for disks is $O(\frac{r^2}{\epsilon^2}Q_p\log\frac{r^2}{\epsilon^2}|K_p|_{\max}+Q_d)$, since $N_d\!=\!\sum_{p=1}^{N_{\text{PRF}}}N_{d,p}\!=\!O(\sum_{p=1}^{N_{\text{PRF}}}\frac{r^2}{\epsilon^2}|K_p|)\!=\!O(\frac{r^2}{\epsilon^2}Q_p)$.
By the same token, if WGD or any sub-index is used for a disk selection heuristic, the complexity $O(N_d\log N_d+Q_d)$ becomes $O(\frac{r^2}{\epsilon^2}Q_p\log\frac{r^2}{\epsilon^2}Q_p+Q_d)$.
\end{proof}
\end{Prop}

For other operations, the same complexity analysis in \ref{sec:complex_bfs} to \ref{sec:complex_mls} can be applied, except some notions are changed from $K_p$ and $Q_p$ to $T_d$ and $Q_d$, respectively.
Therefore, the time complexities of the pulse interleaving algorithm for SDBF using different data structures are arranged as the following lemma.

\begin{Lem}\label{lem4}
The time complexities of the pulse interleaving algorithm for SDBF are summarized as follows:
\begin{enumerate}
\item brute force search: $O(N_t(N_{\text{PRF}}+N_{\text{intlv}}|K_d|_{\max})+Q_d+\frac{r^2}{\epsilon^2}Q_p\log\frac{r^2}{\epsilon^2}|K_p|_{\max})$
\item sorted task lists: $O(N_tN_{\text{PRF}}+Q_dN_{\text{intlv}}^2\log|K_d|_{\max}+\frac{r^2}{\epsilon^2}Q_p\log\frac{r^2}{\epsilon^2}|K_p|_{\max})$
\item orthogonal range trees: $O(N_tN_{\text{PRF}}+Q_d(\log|K_d|_{\max}+\log^2N_{\text{intlv}})+N_tN_{\text{intlv}}\log^2N_{\text{intlv}}+\frac{r^2}{\epsilon^2}Q_p\log\frac{r^2}{\epsilon^2}|K_p|_{\max})$.
\end{enumerate}
The last term $\frac{r^2}{\epsilon^2}Q_p\log\frac{r^2}{\epsilon^2}|K_p|_{\max}$ of each expression is changed to $\frac{r^2}{\epsilon^2}Q_p\log\frac{r^2}{\epsilon^2}Q_p$ if WGD or any sub-index of a disk selection heuristic is used.
\begin{proof}
The time complexities of the calculations for the availabilities of all task-PRF pairs ($O(N_tN_{\text{PRF}})$),  the merge of the schedule ($Q_d$), and the preprocessing for the disks in Proposition \ref{prop8} are common for the three different data structures.
In the algorithm for SDBF, the data structures for the task selection are required for each disk rather than each PRF.
For the brute force search, the time complexity of the task deletion is $Q_d$ and the complexity of a call of {\alg BI} from {\alg HIED} is $N_{\text{intlv}}|K_d|_{\max}$ (refer to Proposition \ref{prop3} and Lemma \ref{lem1}).
For the multiple sorted task lists, the time complexity for the preprocessing is $O(Q_dN_{\text{intlv}}^2\log|K_d|_{\max})$, which dominates the complexities of other operations, such as the task deletion ($O(Q_dN_{\text{intlv}}^2)$) and the calls of {\alg BI} ($O(N_tN_{\text{intlv}}^2)$) (refer to Lemma \ref{lem2}).
For the orthogonal range trees, the time complexity for the preprocessing, the task deletion, and the calls of {\alg BI} are $O(Q_d(\log|K_d|_{\max}+\log^2N_{\text{intlv}}))$, $O(Q_d\log^2N_{\text{intlv}})$, and $O(N_tN_{\text{intlv}}\log^2N_{\text{intlv}})$, respectively (refer to Lemma \ref{lem3}).
Then the lemma is concluded by Proposition \ref{prop8}.
\end{proof}
\end{Lem}

As in Theorem \ref{thm1}, the time complexities in Lemma \ref{lem4} can be simplified with constant parameters $N_{\text{PRF}}$, $N_{\text{intlv}}$, $r$, and $\epsilon$ and a scalable number of targets.

\begin{Thm}\label{thm2}
The time complexity of the pulse interleaving algorithm for SDBF using multiple sorted task lists or orthogonal range trees is $O(N_t^2\log N_t)$ if $N_{\text{PRF}}$, $N_{\text{intlv}}$, $r$, and $\epsilon$ are constant.
\begin{proof}
Since $|K_p|_{\max}\leq N_t$, $|K_d|_{\max}\leq N_t$, and $N_d=O(\frac{r^2}{\epsilon^2}Q_p)$, if $N_{\text{PRF}}$, $N_{\text{intlv}}$, $r$, and $\epsilon$ are constant, then $O(Q_p)\leq O(N_tN_{\text{PRF}})=O(N_t)$, $O(Q_d)\leq O(N_tN_d)=O(N_tQ_p)=O(N_t^2)$, and the complexities in Lemma \ref{lem4} are reduced to $O(N_t^2\log N_t)$.
\end{proof}
\end{Thm}

\section{Conclusions}\label{sec:con}

This paper has formulated the interleaved pulse scheduling problem using multiple simultaneous received beams (MSRB) for multiple target tracking in a pulse Doppler phased array radar (PAR) by an integer program.
The problem formulation is valid for both element and subarray level digital beamforming (DBF) architectures, and heuristic pulse interleaving algorithms are presented for the problems of the different DBF levels.
The complexity analysis show that the heuristic pulse interleaving algorithms can be executed in a practical computation time.

\section*{Acknowledgment}

This work was supported by Basic Science Research Program through the National Research Foundation of Korea (NRF) funded by the Ministry of Science, ICT and Future Planning (NRF-2013R1A1A1008693).

\bibliography{COR_PIMSRB}
\bibliographystyle{IEEEtran}

\end{document}